\documentclass[aps,prd,a4paper,onecolumn,amsmath,showpacs,superscriptaddress,nofootinbib,preprintnumbers,notitlepage]{revtex4-1}
 
\usepackage{verbatim}
\usepackage[T1]{fontenc}
\usepackage[utf8]{inputenc}
\usepackage[american]{babel}
\usepackage{epsfig}
\usepackage{graphicx,subcaption,caption}
\usepackage{pgfplots}
\usepackage{booktabs}
\usepackage{multirow}
\usepackage{dcolumn}
\usepackage{amsmath}
\usepackage{mathtools}
\usepackage{amsfonts}
\usepackage{amssymb}
\usepackage{epstopdf}
\usepackage{bm}
\usepackage{siunitx}
\usepackage{braket}
\usepackage{enumitem}
\usepackage{soul}
\usepackage{color}
\usepackage{transparent}
\usepackage{pifont}
\usepackage[font={small}]{caption}
\usepackage{enumitem}

\definecolor{navyblue}{rgb}{0.0, 0.0, 0.5}
\definecolor{royalblue}{rgb}{0.25, 0.41, 0.88}
\definecolor{cadmiumgreen}{rgb}{0.0, 0.42, 0.24}
\definecolor{blue-violet}{rgb}{0.54, 0.17, 0.89}
\definecolor{darkviolet}{rgb}{0.58, 0.0, 0.83}
\definecolor{orange(colorwheel)}{rgb}{1.0, 0.5, 0.0}

\usepackage{hyperref}
\hypersetup{
    colorlinks=true, % false: boxed links; true: colored links
    linkcolor=royalblue, % color of internal links (change box color with linkbordercolor)
    citecolor=magenta}
\newlist{todolist}{itemize}{2}

\newcommand\be{\begin{equation}}
\newcommand\ee{\end{equation}}

\newcommand\bea{\begin{eqnarray}}
\newcommand\eea{\end{eqnarray}}

\usepackage{booktabs}
\usepackage{multirow}
\usepackage{dcolumn}
\usepackage{colortbl}

\definecolor{magenta(process)}{rgb}{1.0, 0.0, 0.56}

\definecolor{darkspringgreen}{rgb}{0.09, 0.45, 0.27}

\definecolor{royalblue(web)}{rgb}{0.25, 0.41, 0.88}

\begin{document}

\title{Traversable wormhole in logarithmic $f(R)$ gravity by various shape and redshift functions}

\author{Jafar Sadeghi}
\email{pouriya@ipm.ir}
\affiliation{Department of Physics, University of Mazandaran, P. O. Box 47416-95447, Babolsar, Iran}
\affiliation{School of Physics, Damghan University, P. O. Box 3671641167, Damghan, Iran}

\author{Mehdi Shokri}
\email{mehdishokriphysics@gmail.com}
\affiliation{School of Physics, Damghan University, P. O. Box 3671641167, Damghan, Iran}

\author{Saeed Noori Gashti}
\email{saeed.noorigashti@stu.umz.ac.ir}
\affiliation{Department of Physics, University of Mazandaran, P. O. Box 47416-95447, Babolsar, Iran}

\author{Behnam Pourhassan}
\email{b.pourhassan@du.ac.ir}
\affiliation{School of Physics, Damghan University, P. O. Box 3671641167, Damghan, Iran}
\affiliation{Iran Science Elites Federation, Tehran, Iran}
\affiliation{Canadian Quantum Research Center 204-3002 32 Avenue Vernon, British Columbia V1T 2L7, Canada}

\author{Prabir Rudra}
\email{prudra.math@gmail.com}
\affiliation{Department of Mathematics, Asutosh College, Kolkata-700 026, India}

\preprint{}
\begin{abstract}
We study the traversable wormhole solutions for a logarithmic corrected $f(R)$ model by considering two different statements of shape $b(r)$ and redshift $\Phi(r)$ functions. We calculate the parameters of the model including energy density $\rho$, tangential pressure $P_{t}$ and radial pressure $P_{r}$ for the corresponding forms of the functions. Then, we investigate different energy conditions such as null energy condition, weak energy condition, dominant energy condition and strong energy condition for our considered cases. Finally, we explain the satisfactory conditions of energy of the models by related plots.
\end{abstract}

\maketitle
\section{Introduction}
As hypothetical geometric objects, wormholes can solve Einstein equations by tolerating the important condition, namely the violation of null energy conditions (NEC). Wormholes have a tubular structure that is asymptotically flat on both sides. This structure creates a geometry that connects two distinct points of the same space-time or two different universes. The wormholes structure was first introduced by Flamm \cite{1}, but his solutions suffered some stability problems. Later, Einstein and Rosen studied a similar structure of the wormhole and introduced a concept called the Einstein-Rosen Bridge \cite{2}. These hypothetical objects lead to different static and non-static in proportion to the constant or variable radius of the wormhole throat. Wormholes have been investigated from different aspects and conditions \cite{3,4,5,6,7,8,9,10,11,12,13,14,15,17,18,19,20,21,22,23,24,25,26,27,28}. Shortly afterward, the Kerr wormholes were studied, which have a quantum structure and connect different points of space on the Planck scale. Despite the mentioned features, these wormholes were not traversable \cite{29}. As a further development, later Thorne and Morris \cite{3} specially laid the structure of wormholes with some new concepts such as the throat. They used general relativity concepts and found a way to journey as a wonderful and laudable work. It is worth noting that the existence of wormholes requires ''exotic'' mater that violate the null energy condition \cite{17,31}. As we know, there is a lack of the matter to tolerate the wormhole geometry. This has raised a susceptible question of whether these modified theories of gravity, which are used in many other concepts such as the expansion of the universe, could also answer the structure of the wormhole. One of the most successful modified theories of gravity is $f(R)$ theory which considers a general form of the Ricci scalar $R$ instead of $R$ in the form of the Einstein-Hilbert action \cite{32,33,34,35, 36,37,38,39,40,41,42,43,44,45,46,47,48,49,51,52,53,54,55,56,57,58,59,60,61,
62,64,65,66}. Besides the successes of the model for Dark Energy (DE) and also other cosmological scenarios, $f(R)$ gravity has been engaged in some works in order to study the wormholes solutions \cite{n1,n2}. Also, one can the wormholes analysis in $f(R,T)$ gravity in refs \cite{n3,n4,n5}. Such studies have shown that modifications derived from higher curvatures will be responsible for violating energy conditions using different forms of redshift and various shape functions. Moreover, the static wormholes in higher dimensions have also been investigated \cite{77,78}. On the other hand, the studies of wormholes have been performed in energy density and radial pressure and tangential pressure proportional to the wormhole mouth. In \cite{79,80}, the authors studied traversable wormholes according to the FLRW model and evaluated the effects of additional dimensions. Also, different types of wormholes in the Finslerian structure of space-time have also been studied. Other types of shape and redshift functions  and their field equations are investigated by \cite{81,82}. In \cite{85}, Novikov reviewed and categorized the different types of wormholes. They determined the properties associated with wormholes and their relations to black holes \cite{86,87,88,89,90,91,92,93,94,95,96}. The structure of spherically symmetric static wormholes has been studied in \cite{97}.

In this paper, we study the issue of wormholes in the context of the logarithmic corrected $f(R)$ model by different forms of shape and redshift functions. Such gravitational theories with modifications in geometry tell us that traversable wormholes can be occurred without any exotic matter. Above discussions motivate us to arrange the paper as follows.  
In \S II, we describe the field equations and wormhole geometry in the context of $f(R)$ gravity. In \S III, we carry out the analysis for the logarithmic corrected $f(R)$ model $R+\alpha R^{2}+\beta R^{n}+\gamma R^{2}\log \gamma R$ for different forms of shape $b(r)$ and redshift $\Phi(r)$ functions. Then, we investigate the satisfactory conditions of energy by the parameters $\rho$, $P_{r}$ and $P_{t}$ of the models. In \S IV, we conclude the analysis of the models and draw the possible outlooks.
\section{Wormholes in $f(R)$ theory}
To describe the wormhole structure, we start with static and symmetric spherical metric
\begin{equation}
ds^{2}=-e^{2\Phi(r)}dt^{2}+\frac{dr^{2}}{1-\frac{b(r)}{r}}+r^{2}(d\theta^{2}+\sin^{2}\theta d\phi^{2}),
\label{1}
\end{equation}
where $b(r)$ and $\Phi(r)$ are shape and redshift functions, respectively. On the other hands, the form of action in $f(R)$ gravity takes the form
\begin{equation}
S=\frac{1}{2\kappa}\int(f(R)+L_{m})\sqrt{-g}d^{4}x,
\label{2}
\end{equation}
where $g$ is the determinant of the metric $g_{\mu\nu}$, $R=g^{\mu\nu}R_{\mu\nu}$ is the Ricci scalar as gravitational sector and $\mathcal{L}_{M}$ is the Lagrangian of matter filling the universe as perfect fluid. Also, here we suppose $\kappa^{2}\equiv8\pi G=1$. By varying the action (\ref{2}) respect to the metric, the Einstein field equation is given by
\begin{equation}
FR_{\mu\nu}-\frac{1}{2}fg_{\mu\nu}-\nabla_{\mu}\nabla_{\nu}F+\Box Fg_{\mu\nu}=T^{m}_{\mu\nu},
\label{3}
\end{equation}
where $F=\frac{df}{dr}$. Multiplying by $g^{\mu\nu}$, we have
\begin{equation}
FR-2f+3\Box F=T,
\label{4}
\end{equation}
where $T=T^{\mu}_{\mu}$. Combining the Eqs. (\ref{3}) and (\ref{4}), the effective field equations are calculated as 
\begin{equation}
G_{\mu\nu}\equiv R_{\mu\nu}-\frac{1}{2}Rg_{\mu\nu}=T^{eff}_{\mu\nu},\hspace{0.5cm} with \hspace{0.5cm} T^{eff}_{\mu\nu}=T^{c}_{\mu\nu}+\frac{T^{m}_{\mu\nu}}{F},
\label{5}
\end{equation}
where
\begin{equation}
T^{c}_{\mu\nu}=\frac{1}{F}(\nabla_{\mu}\nabla_{\nu}F-\frac{1}{4}g_{\mu\nu}(FR+\Box F+T).
\label{6}
\end{equation}
The energy-momentum tensor $T_{\mu\nu}$ can be introduced as
\begin{equation}
T_{\mu\nu}=(\rho+P_{t})u_{\mu}u_{\nu}-P_{t}g_{\mu\nu}+(P_{r}-P_{t})\chi_{\mu}\chi_{\nu},
\label{7}
\end{equation}
where $u_{\mu}$ is the four-velocity and $\chi_{\mu}$ is the unit space-like vector. Also, $\rho$, $P_{r}$ and $P_{t}$ are energy density, radial pressure measured in the direction of $\chi_{\mu}$ and tangential pressure measured in the orthogonal direction of $\chi_{\mu}$, respectively. Using the metric (\ref{1}) and above expression, the Einstein equation can be solved as
\begin{equation}
\rho=\frac{Fb'(r)}{r^{2}}-(1-\frac{b(r)}{r})F'\Phi'(r)-H,
\label{8}
\end{equation}
\begin{equation}
P_{r}=-\frac{b(r)F}{r^{3}}+2(1-\frac{b(r)}{r})\frac{\Phi'(r)F}{r}-(1-\frac{b(r)}{r})
\{F''+\frac{F'(rb'(r)-b(r))}{2r^{2}(1-\frac{b(r)}{r})}\}+H,
\label{9}
\end{equation}
and
\begin{equation}
P_{t}=\frac{F(b(r)-rb'(r))}{2r^{3}}-\frac{F'}{r}(1-\frac{b(r)}{r})+F(1-\frac{b(r)}{r})(\Phi''(r)-\frac{(rb'(r)-b(r))\Phi'(r)}{2r(r-b)}\\+\Phi'^{2}(r)+\frac{\Phi'(r)}{r})+H,
\label{10}
\end{equation}
where $H=\frac{1}{4}(FR+\Box F+T)$.
Also, prime denotes the derivative with respect to radial coordinate $r$. The equation of state parameter or radial state parameter, mass function and anisotropy parameter are introduced, respectively as
\begin{equation}
\omega=\frac{P_{r}}{\rho},\hspace{1cm}m=\int_{r_{0}}^{r}4\pi r^{2}\rho dr,\hspace{1cm}\Delta=P_{t}-P_{r}.
\label{11}
\end{equation}
$\Delta$ can be expressed in three cases i) $\Delta<0$ for attractive geometry ii) $\Delta>0$ for repulsive geometry iii) $\Delta=0$ for a geometry contains isotropic pressure.
\section{Logarithmic corrected model}
$f(R)$ models without a logarithmic correction are commonly used to study a neutron star with a strong magnetic field and other cosmological implications. Logarithmic modifications are also used for the effects of gluons in non-planar space. For further studying plus mentioned above, you can see \cite{97,98,99,100,101,102,103,104,105}. In order to study wormholes in the context of a logarithmic corrected model, we work with a $f(R)$ model contaminated with the logarithmic term as 
\begin{equation}
f(R)=R+\alpha R^{2}+\beta R^{n}+\gamma R^{2}\log \gamma R,\hspace{1cm}0<n<1,
\label{12}
\end{equation}
where $\alpha$, $\beta$ and $\gamma$ are constant coefficients. In this section, we consider four different statements of shape and redshift functions discussed in detail in \cite{j,22}:
\begin{itemize}
  \item $b(r)=\frac{r}{\exp(r-r_{0})}$, $\Phi(r)=\frac{1}{r}$.
  \item $b(r)=\frac{r}{\exp(r-r_{0})}$, $\Phi(r)=c$.
  \item $b(r)=\frac{r_{0}\log(1+r)}{\log(1+r_{0})}$. $\Phi(r)=\frac{1}{r}$. 
  \item $b(r)=\frac{r_{0}\log(1+r)}{\log(1+r_{0})}$, $\Phi(r)=c$.
\end{itemize} 
We calculate the parameters $\rho$, $P_{r}$ and $P_{t}$ of all cases and then we examine the satisfactory conditions of energy by the corresponding energy conditions:
\begin{itemize}
  \item Null Energy Condition (NEC) for $\rho+P_{r}\geq0$ and $\rho+P_{t}\geq0$ .
  \item Weak Energy Condition (WEC) for $\rho \geq0$, $\rho+P_{r}\geq0$ and $\rho+P_{t}\geq0$.
  \item Strong Energy Condition (SEC) for $\rho+P_{r}\geq0$ , $\rho+P_{t}\geq0$ and $\rho+P_{r}+2P_{t}\geq0$.
  \item Dominant Energy Condition (DEC) for $\rho>0$, $\rho-|P_{r}|>0$, $\rho-|P_{t}|>0$ and $\rho+P_{r}+2P_{t}\geq0$. 
\end{itemize} 
\subsection{Case i}
\begin{figure*}[!hbtp]
    \centering
	\includegraphics[width=.40\textwidth,keepaspectratio]{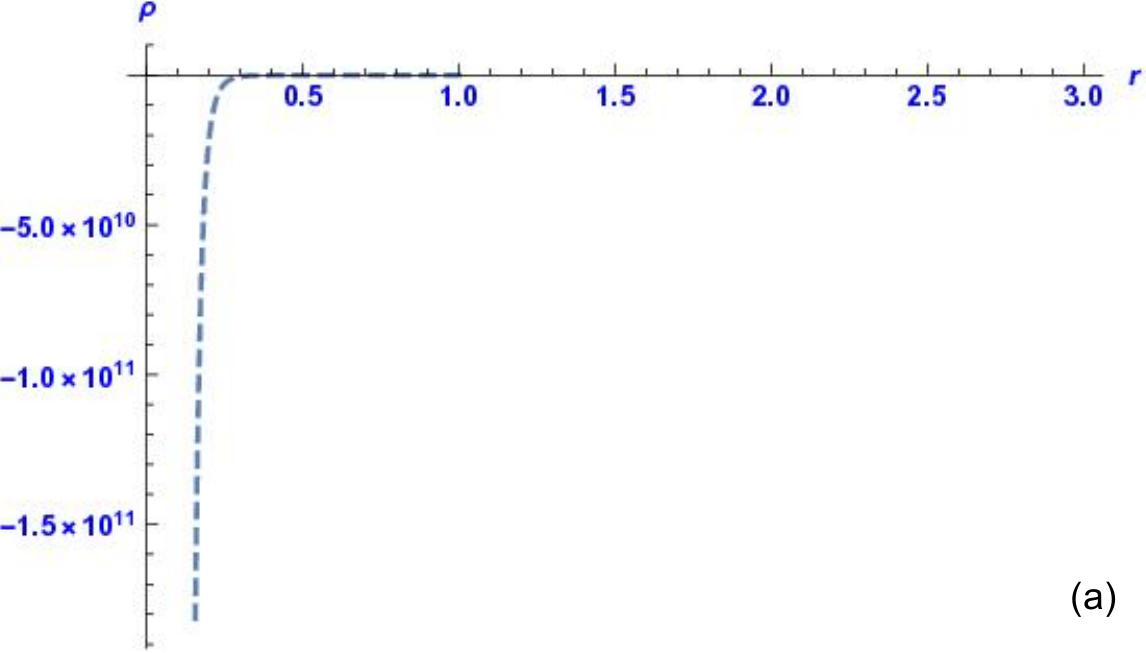}
	\hspace{1cm}
	\includegraphics[width=.40\textwidth,keepaspectratio]{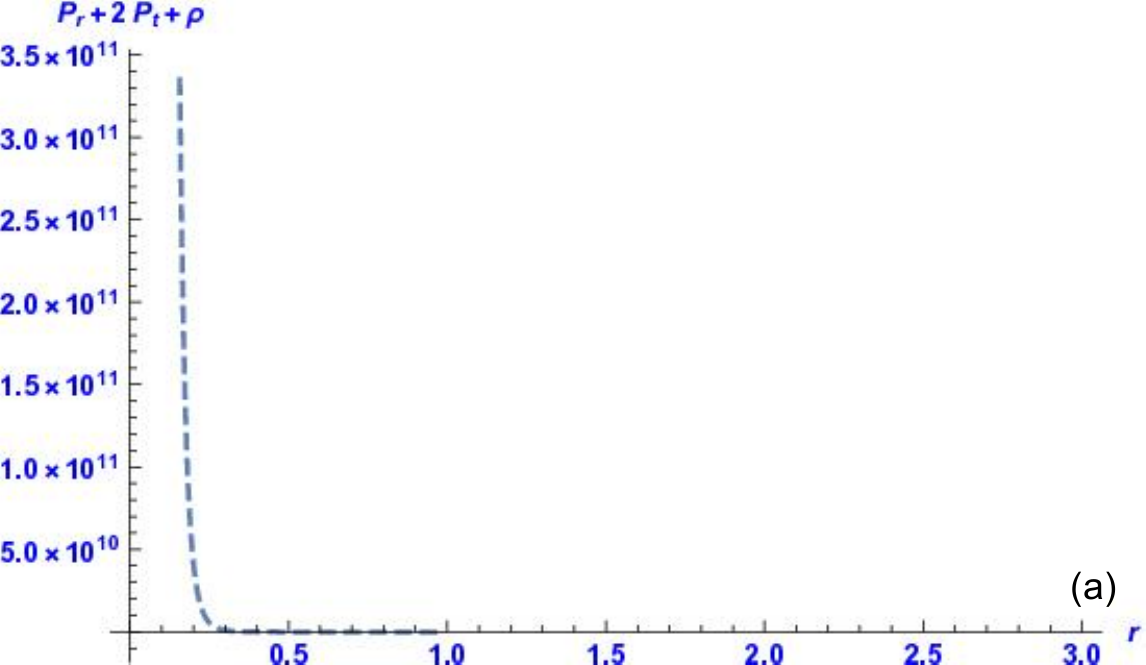}
	\vspace{1cm}
	\includegraphics[width=.40\textwidth,keepaspectratio]{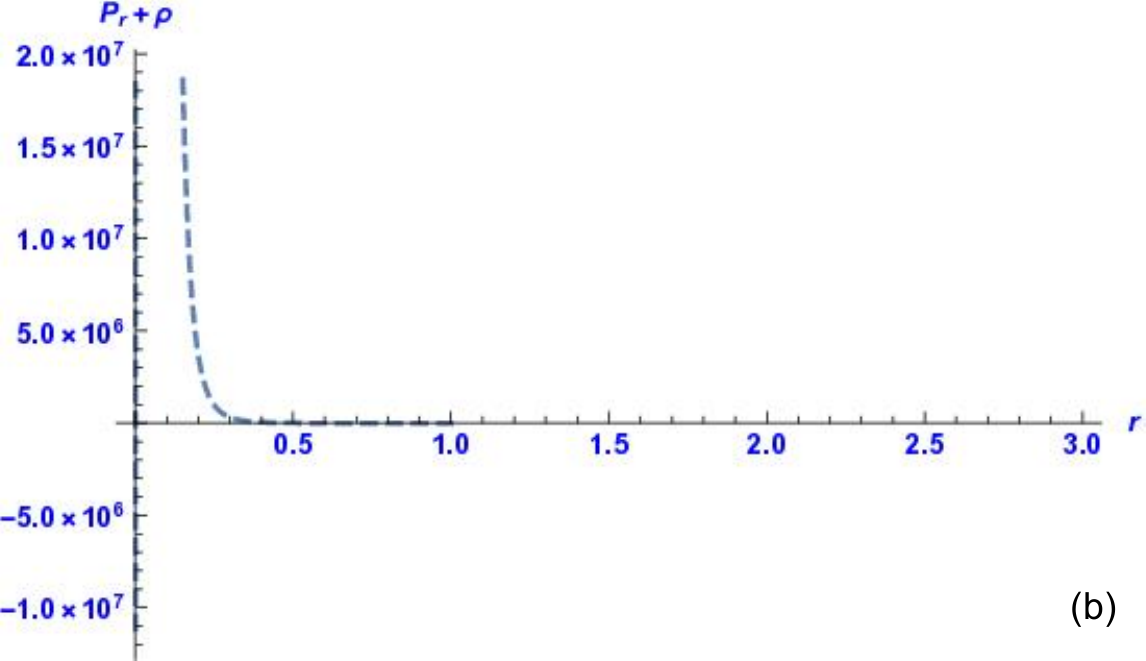}
	\hspace{1cm}
	\includegraphics[width=.40\textwidth,keepaspectratio]{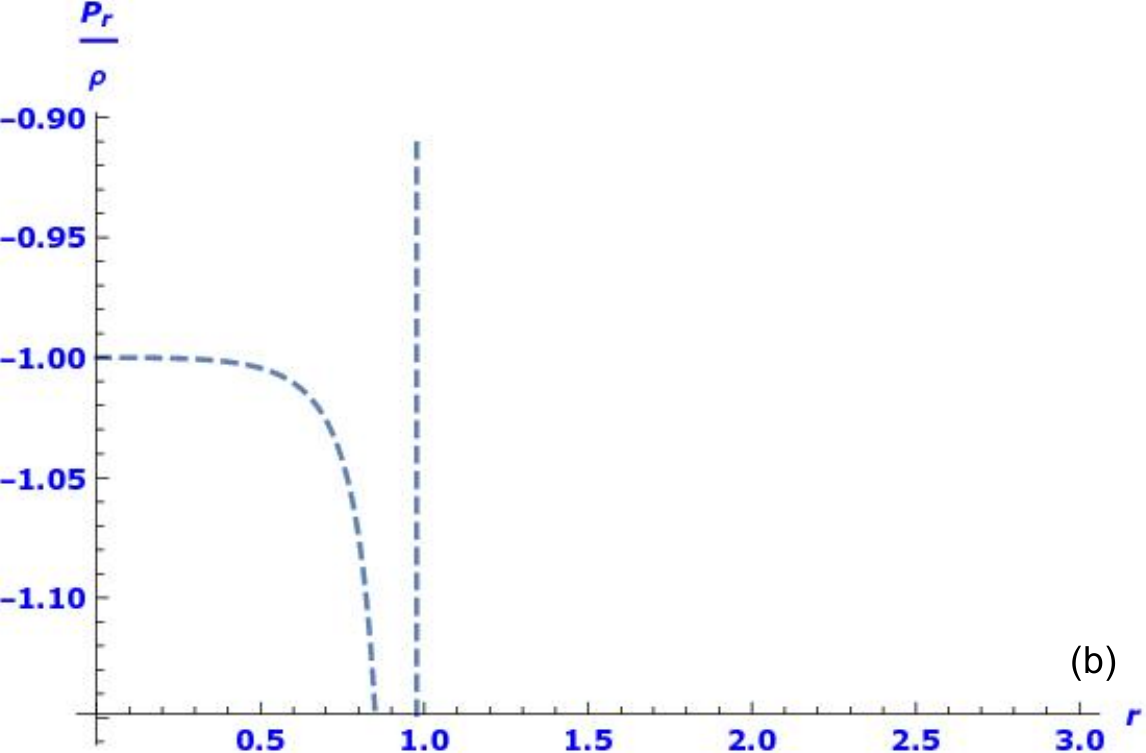}
	\includegraphics[width=.40\textwidth,keepaspectratio]{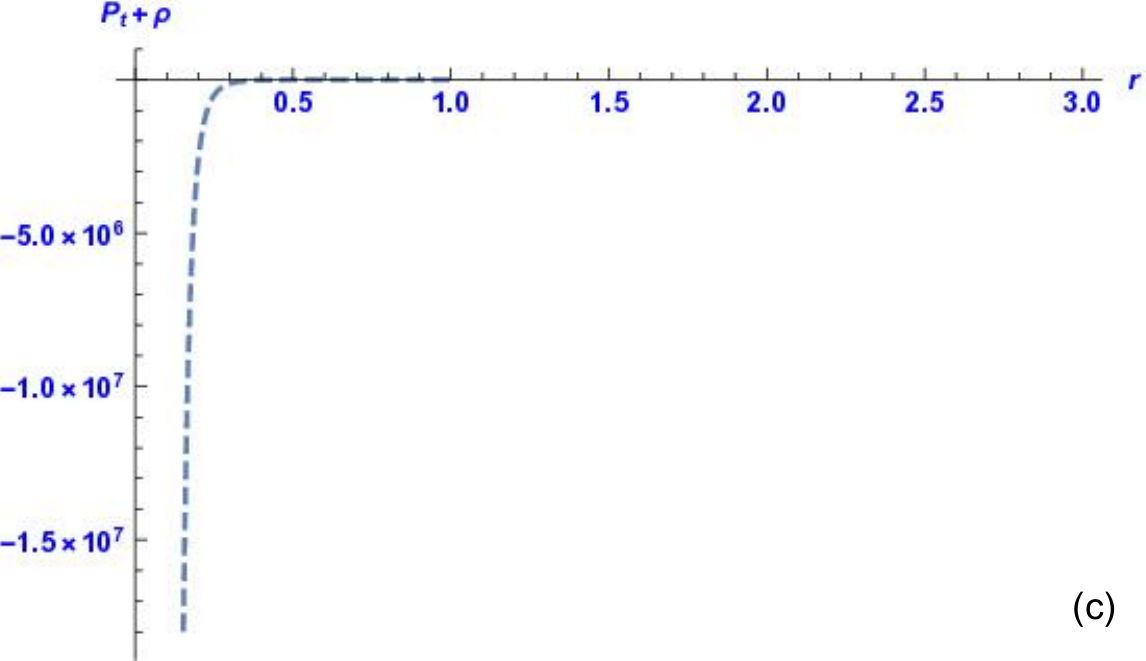}
	\hspace{1cm}
	\includegraphics[width=.40\textwidth,keepaspectratio]{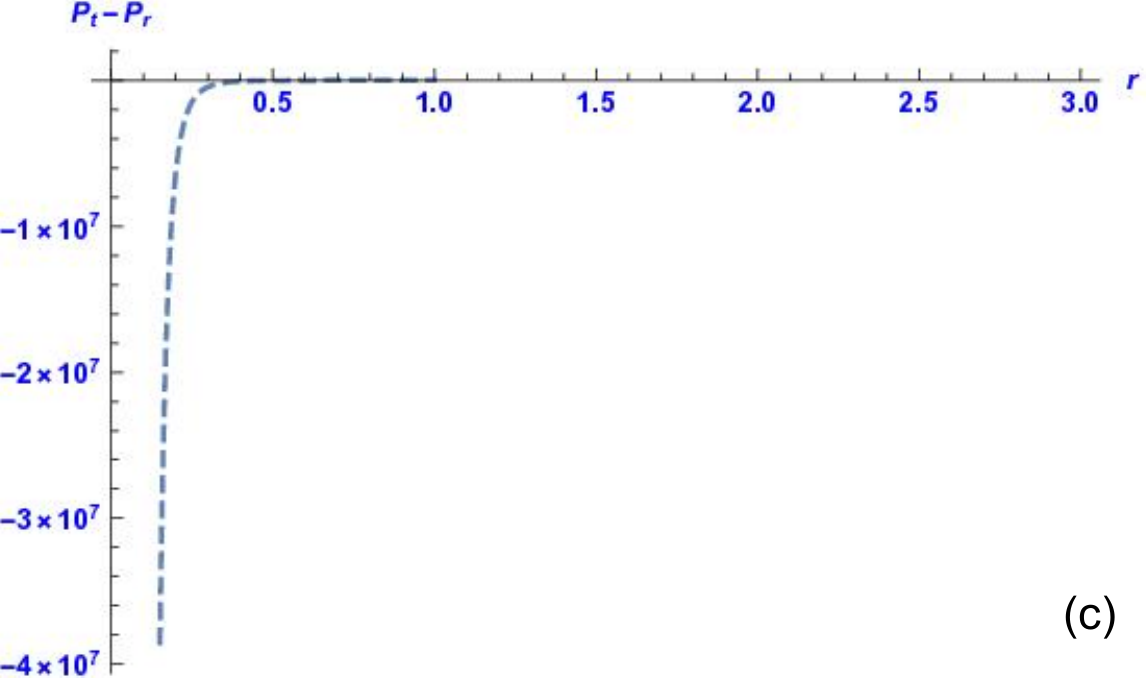}
	\includegraphics[width=.40\textwidth,keepaspectratio]{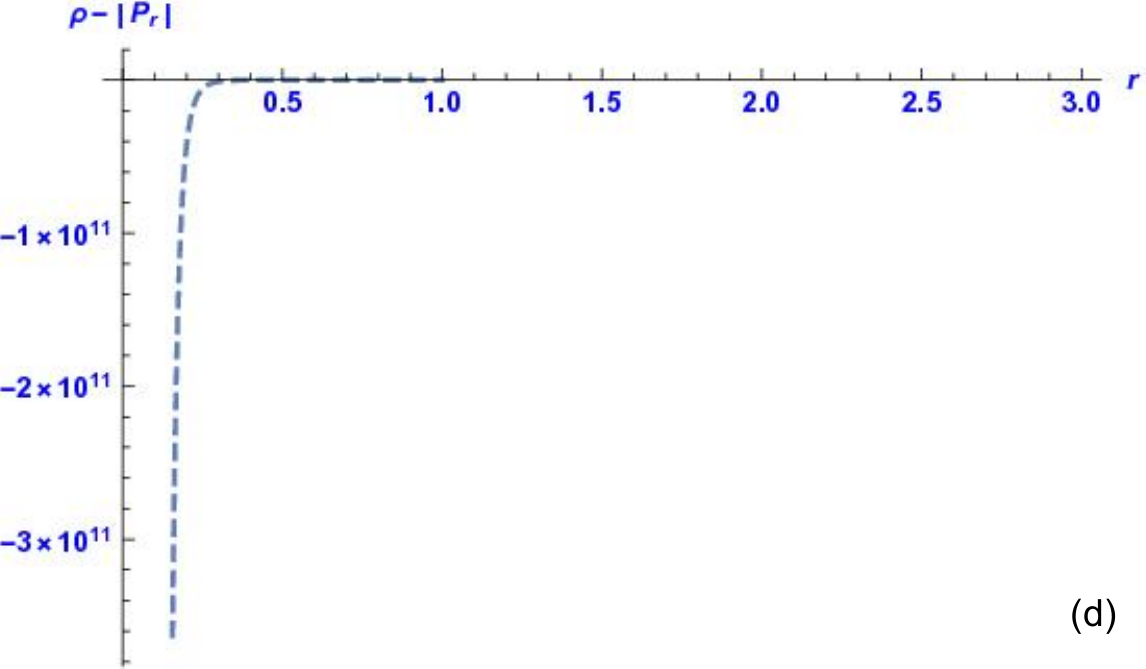}
	\hspace{1cm}
	\includegraphics[width=.40\textwidth,keepaspectratio]{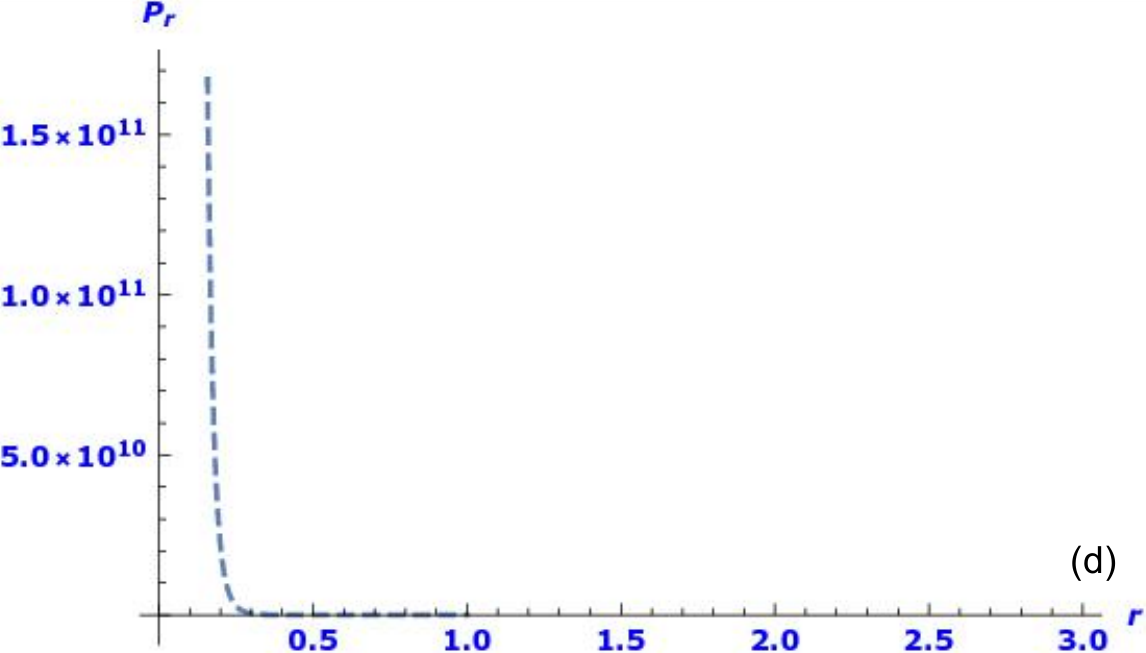}
	\includegraphics[width=.4\textwidth,keepaspectratio]{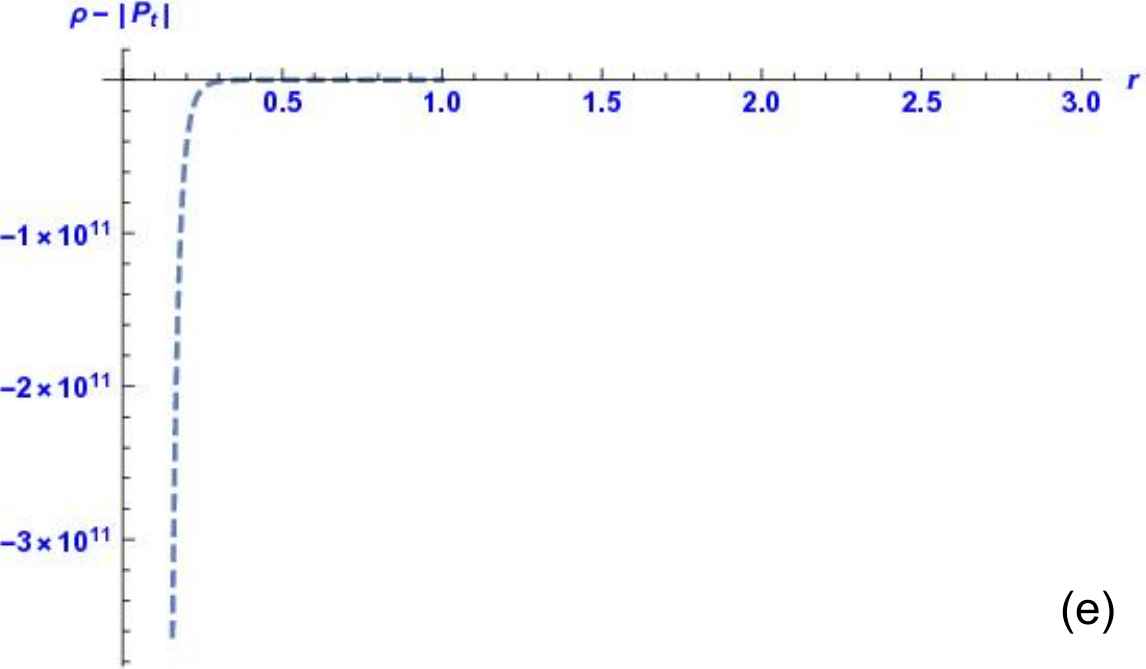}
	\hspace{0.75cm}
	\includegraphics[width=.43\textwidth,keepaspectratio]{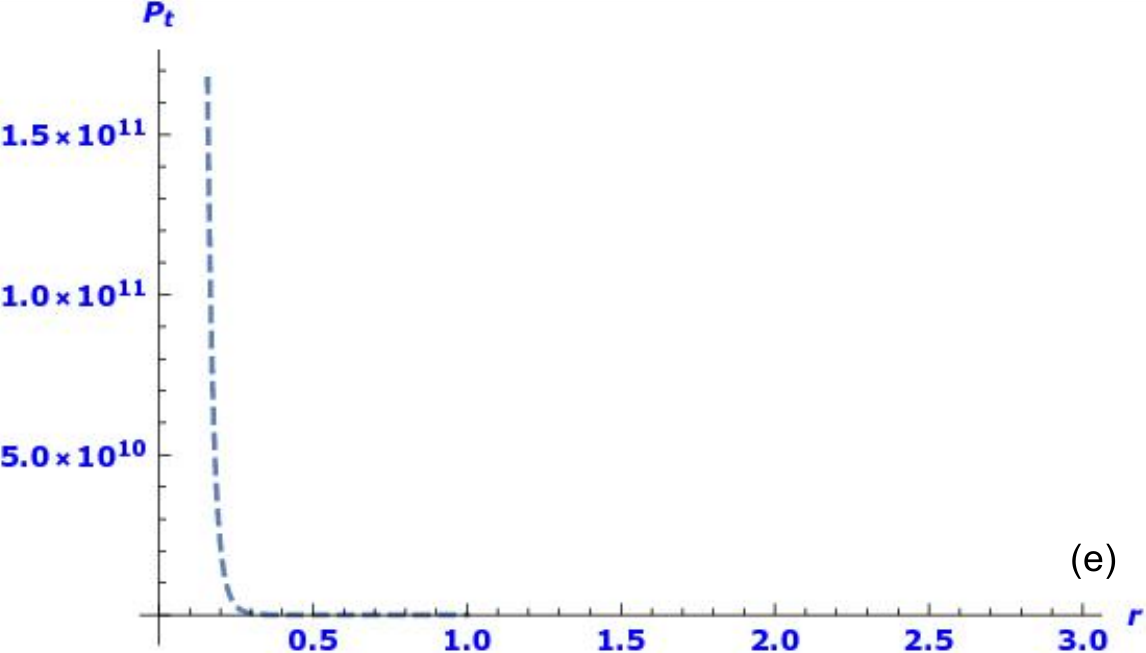}
	\caption{\textit{Left}: $\rho$, $\rho+P_{r}$, $\rho+P_{t}$, $\rho-|P_{r}|$ and $\rho-|P_{t}|$ plotted versus $r$ in the case i. \textit{Right}: $\rho+P_{r}+2P_{t}$, $\frac{P_{r}}{\rho}$, $P_{t}-P_{r}$, $P_{r}$ and $P_{t}$ plotted versus $r$ in the case i.}
	\label{fig1}
\end{figure*}

Let's consider the shape and redshift functions as
\begin{equation}
b(r)=\frac{r}{\exp(r-r_{0})},\hspace{1cm}\Phi(r)=\frac{1}{r}.
\label{13}
\end{equation}
Then, the parameters energy density $\rho$, radius pressure $P_{r}$ and tangential pressure $P_{t}$ are calculated as shown in (\ref{a1}), (\ref{a2}) and (\ref{a5}). In Figure \ref{fig1}, we study the satisfactory situation of different energy conditions for the logarithmic corrected model in the case i (\ref{13}). From the left-side panel (a), we find that WEC is violated since $\rho$ is negative for all values of $r$. Regarding the NEC, the left-side panel (b) reveals that $\rho+P_{r}\geq0$ is fulfilled while $\rho+P_{t}\geq0$ is generally violated as shown in the left-side panel (c). Consequently, case i is not able to satisfy the NEC. From the left-side panels (d) and (e), one can find that the DEC is violated since both corresponding conditions $\rho-|P_{r}|>0$ and $\rho-|P_{t}|>0$ are violated for the model. Also, the right-side panel (a) shows the violation of the SEC condition because $\rho+P_{t}\geq0$ is violated for the model. Moreover, we study two important parameters $w$ and $\Delta$ as the equation of state and anisotropy parameter, respectively. By having in look at the right-side panel (c), we realize that $w$ takes the negative value so that it shows the existence of an exotic matter with a phantom-like behavior. Also, the right-side panel (c) shows an attractive geometry for the wormhole since $\Delta<0$. The right-side panels (d) and (e) presents the behavior of $P{r}$ and $P_{t}$ versus r, respectively. 
\subsection{Case ii}
\begin{figure*}[!hbtp]
    \centering
	\includegraphics[width=.40\textwidth,keepaspectratio]{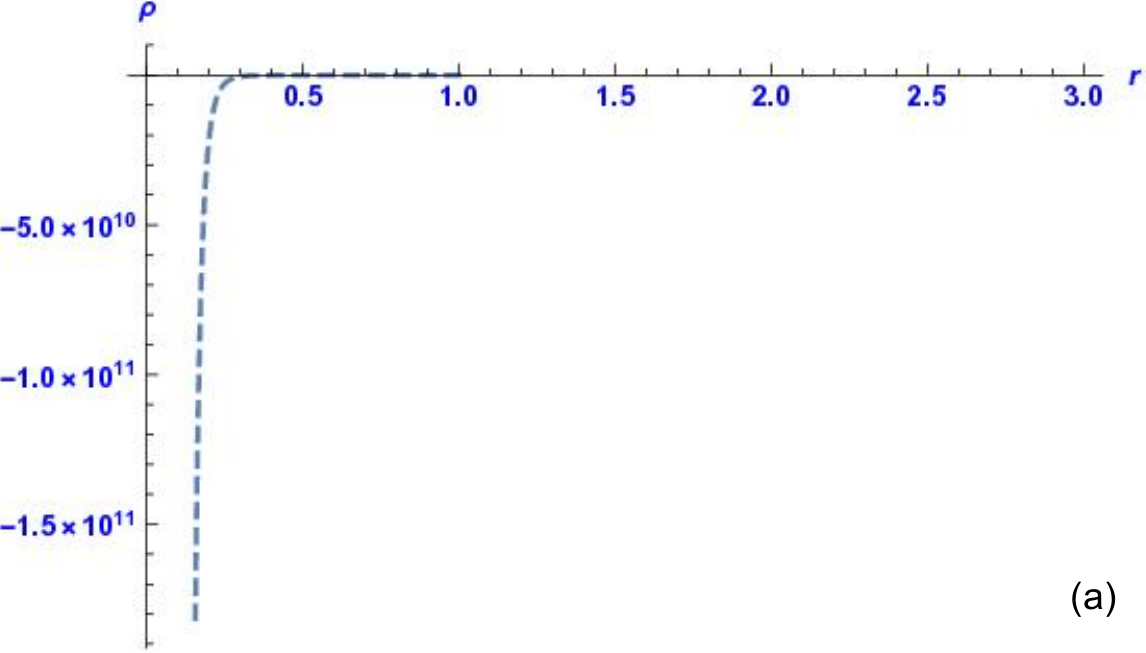}
	\hspace{1cm}
	\includegraphics[width=.40\textwidth,keepaspectratio]{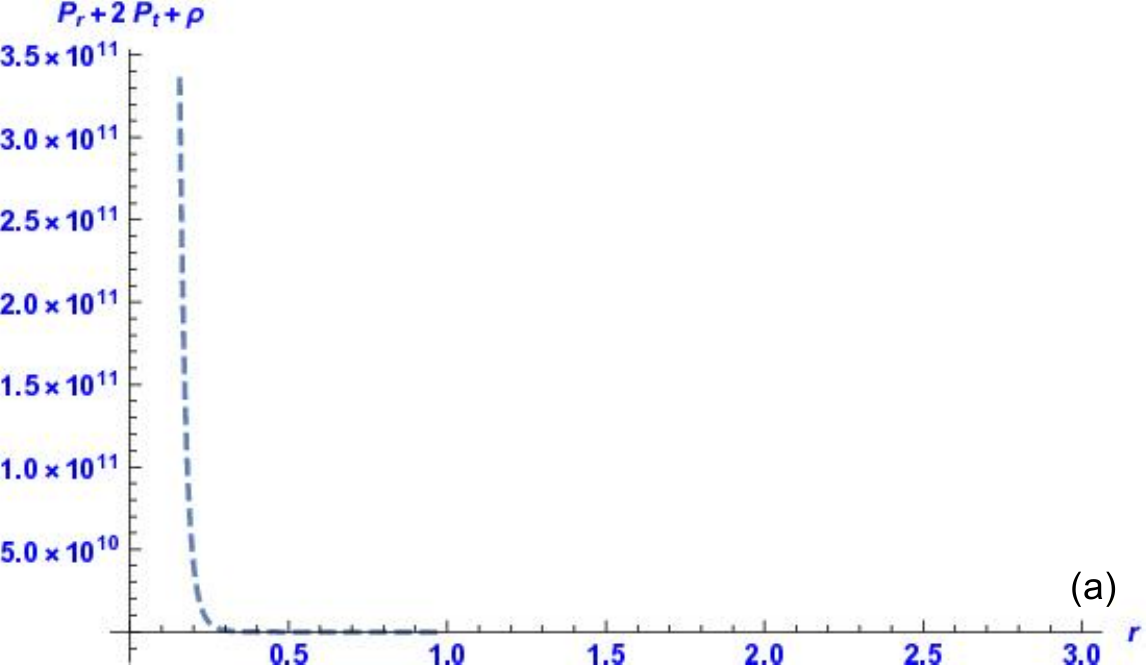}
	\vspace{1cm}
	\includegraphics[width=.40\textwidth,keepaspectratio]{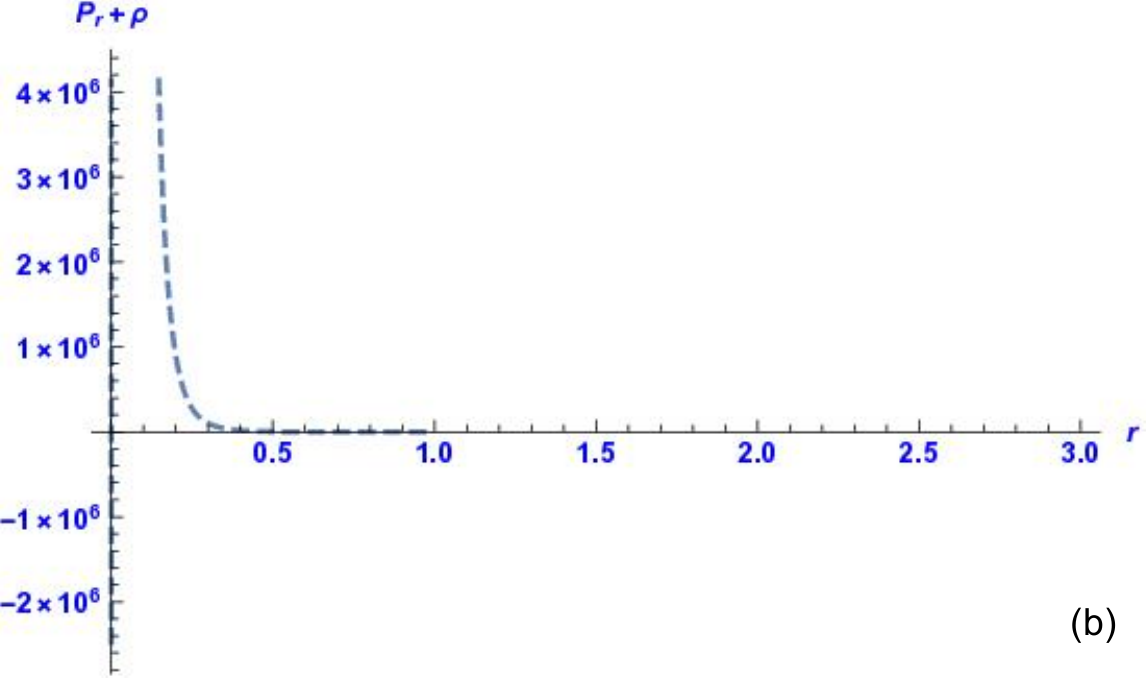}
	\hspace{1cm}
	\includegraphics[width=.40\textwidth,keepaspectratio]{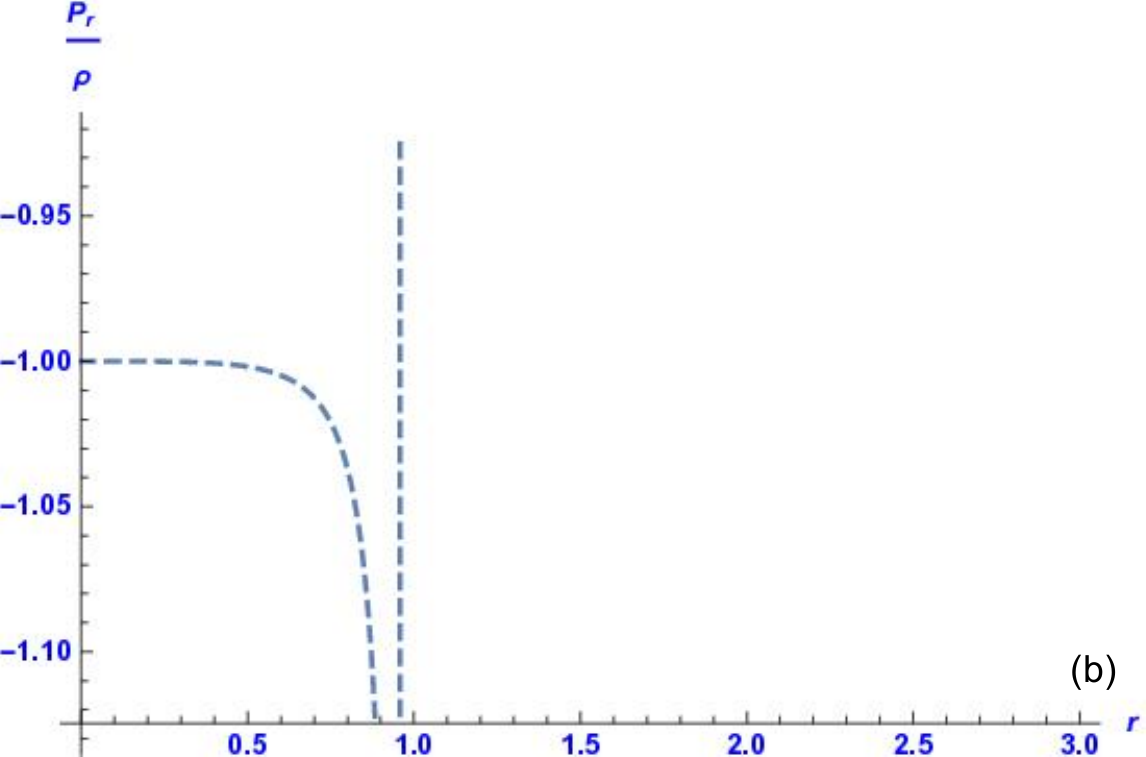}
	\includegraphics[width=.40\textwidth,keepaspectratio]{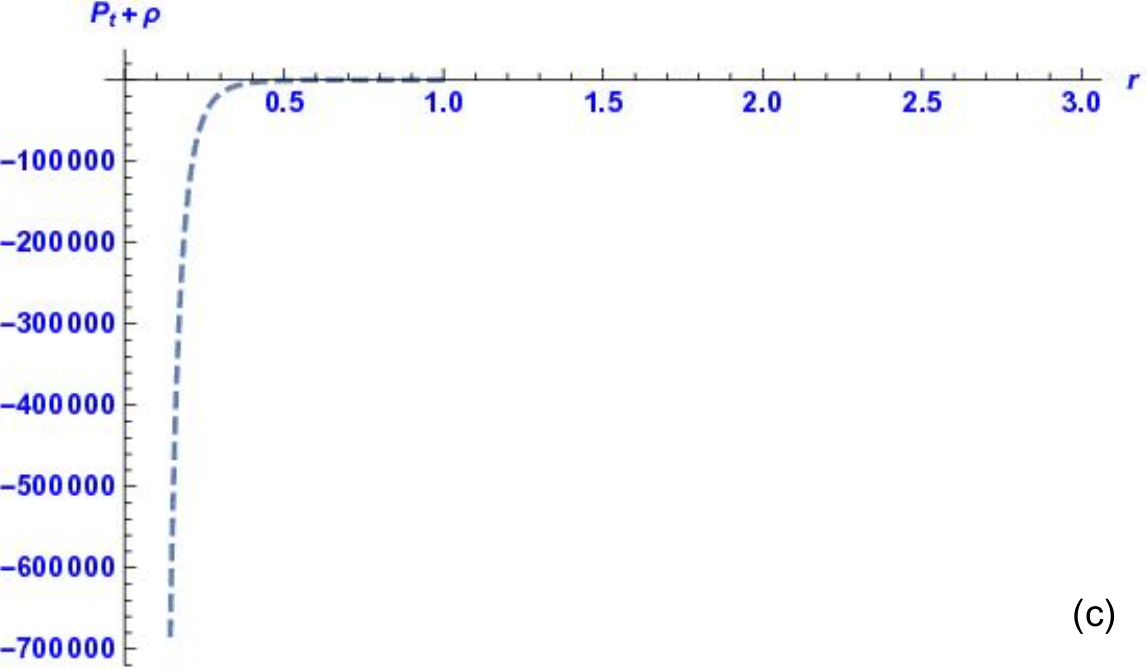}
	\hspace{1cm}
	\includegraphics[width=.40\textwidth,keepaspectratio]{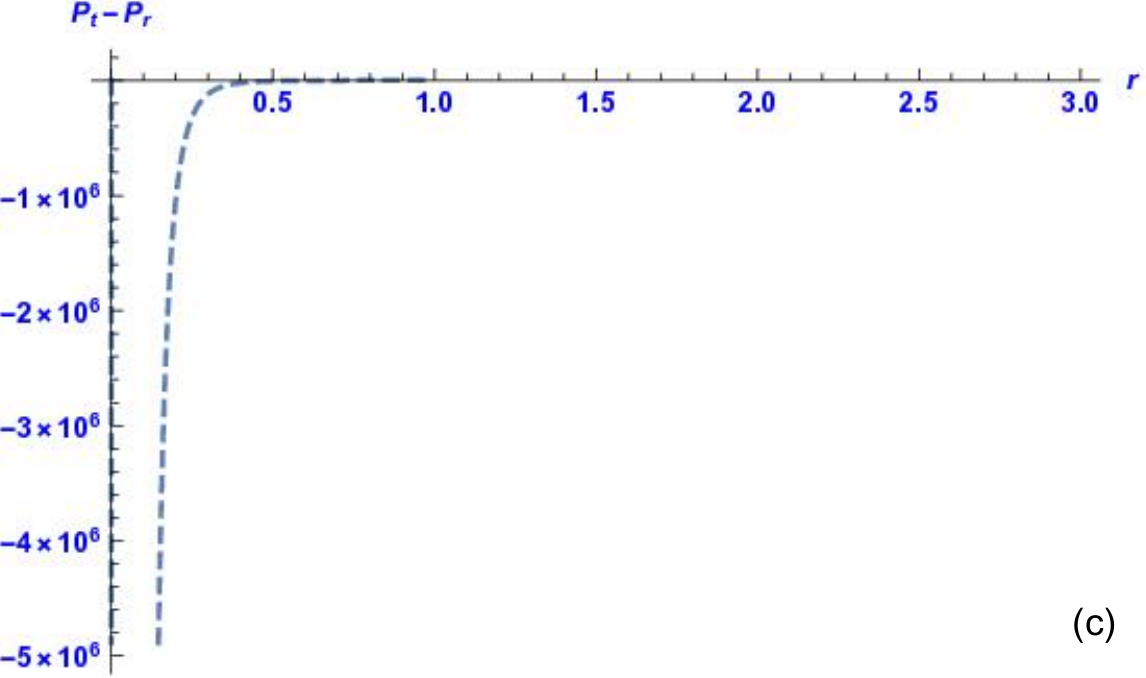}
	\includegraphics[width=.40\textwidth,keepaspectratio]{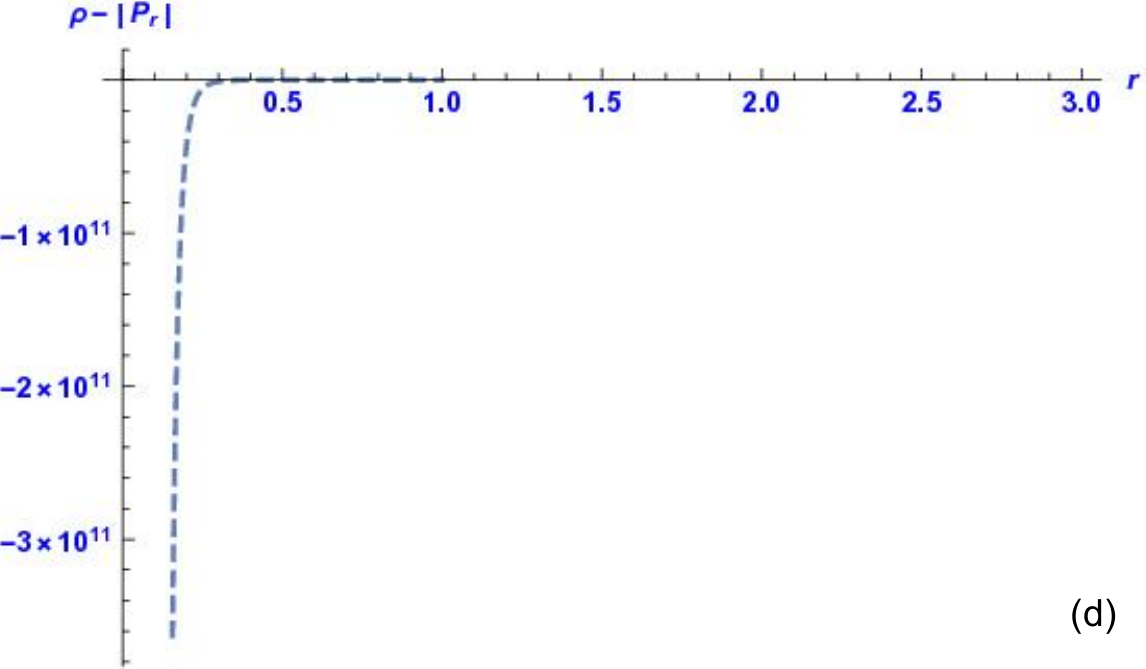}
	\hspace{1cm}
	\includegraphics[width=.40\textwidth,keepaspectratio]{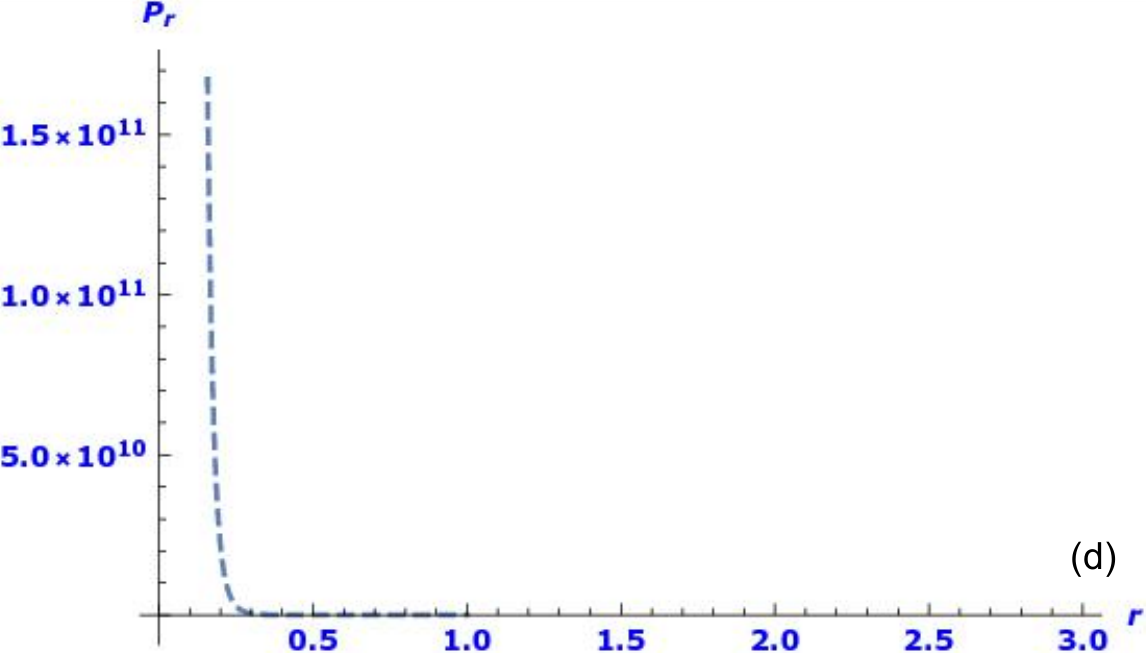}
	\includegraphics[width=.4\textwidth,keepaspectratio]{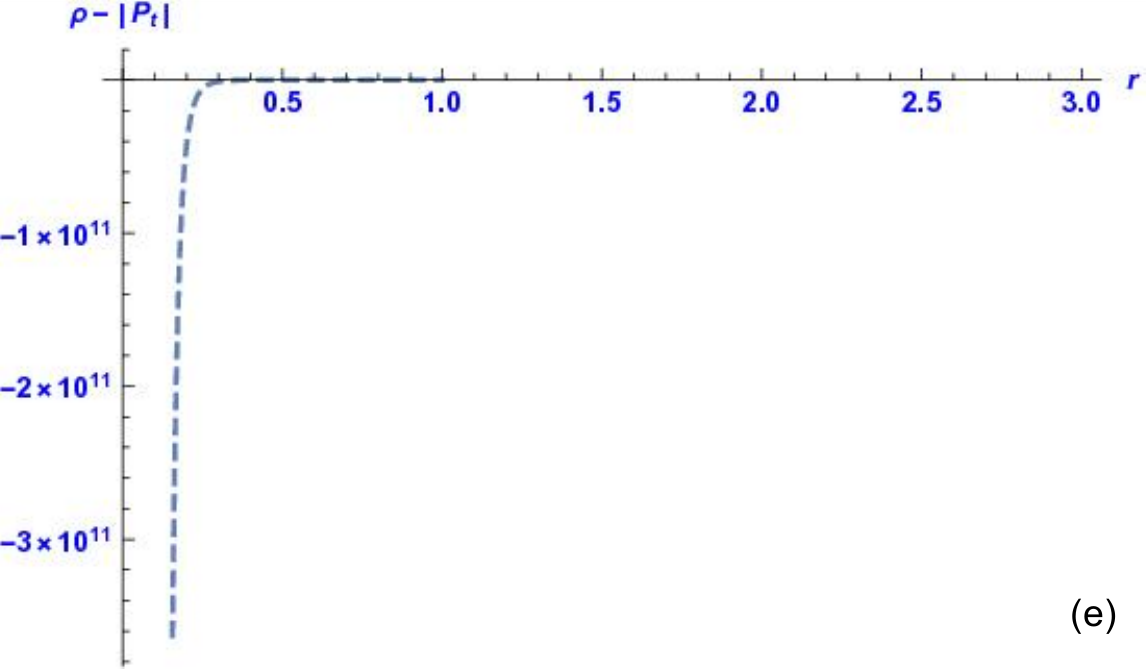}
	\hspace{0.75cm}
	\includegraphics[width=.40\textwidth,keepaspectratio]{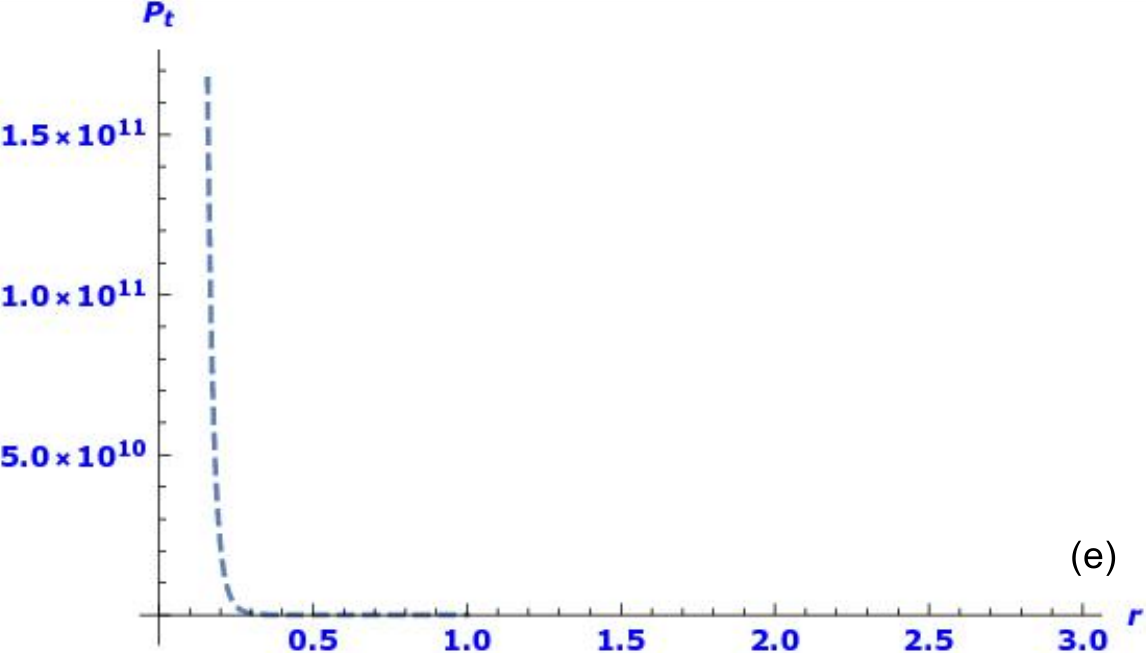}
    \caption{\textit{Left}: $\rho$, $\rho+P_{r}$, $\rho+P_{t}$, $\rho-|P_{r}|$ and $\rho-|P_{t}|$ plotted versus $r$ in the case ii. \textit{Right}: $\rho+P_{r}+2P_{t}$, $\frac{P_{r}}{\rho}$, $P_{t}-P_{r}$, $P_{r}$ and $P_{t}$ plotted versus $r$ in the case ii.}
	\label{fig2}
\end{figure*}

In this case, we define the shape and redshift functions as  
\begin{equation}
b(r)=\frac{r}{\exp(r-r_{0})},\hspace{1cm} \Phi(r)=c 
\label{14}    
\end{equation}
The parameters of the model \textit{i.e.} energy density $\rho$, radius pressure $P_{r}$ and tangential pressure $P_{t}$ are obtained as shown in (\ref{a8}), (\ref{a11}) and (\ref{a12}). In Figure \ref{fig2}, we present the satisfactory situation of different energy conditions for the logarithmic corrected model in case ii (\ref{14}). From the panels of Figure \ref{2}, we obtain the same result with case i for satisfaction situation of different energy conditions but with different values for the related parameters.
\subsection{Case iii}
\begin{figure*}[!hbtp]
    \centering
	\includegraphics[width=.40\textwidth,keepaspectratio]{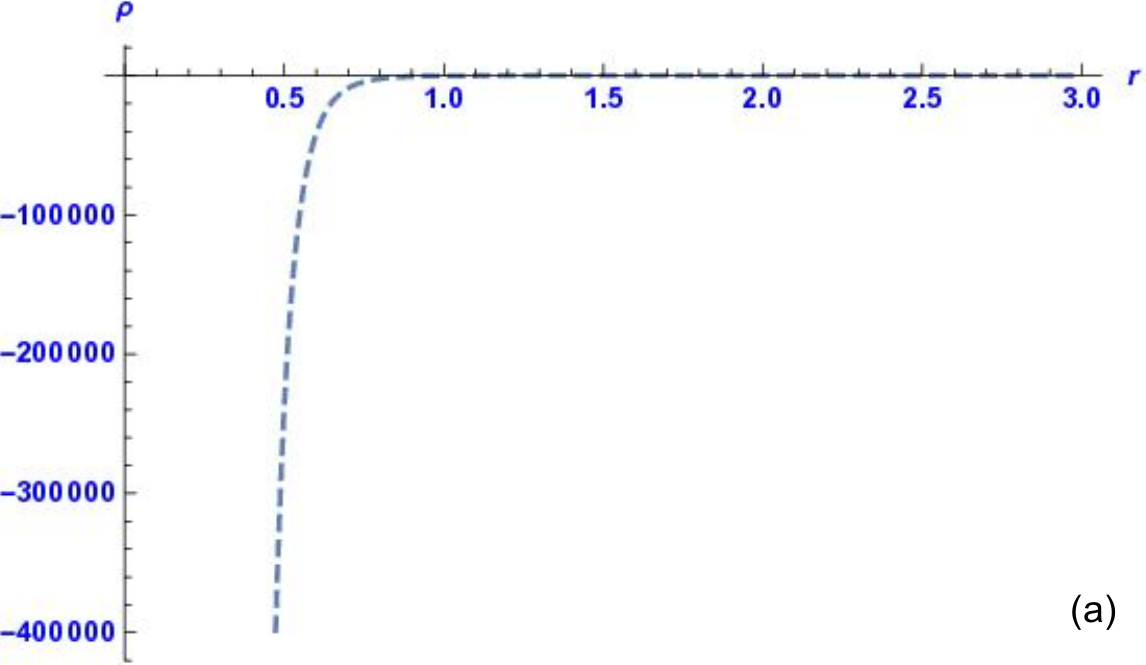}
	\hspace{1cm}
	\includegraphics[width=.40\textwidth,keepaspectratio]{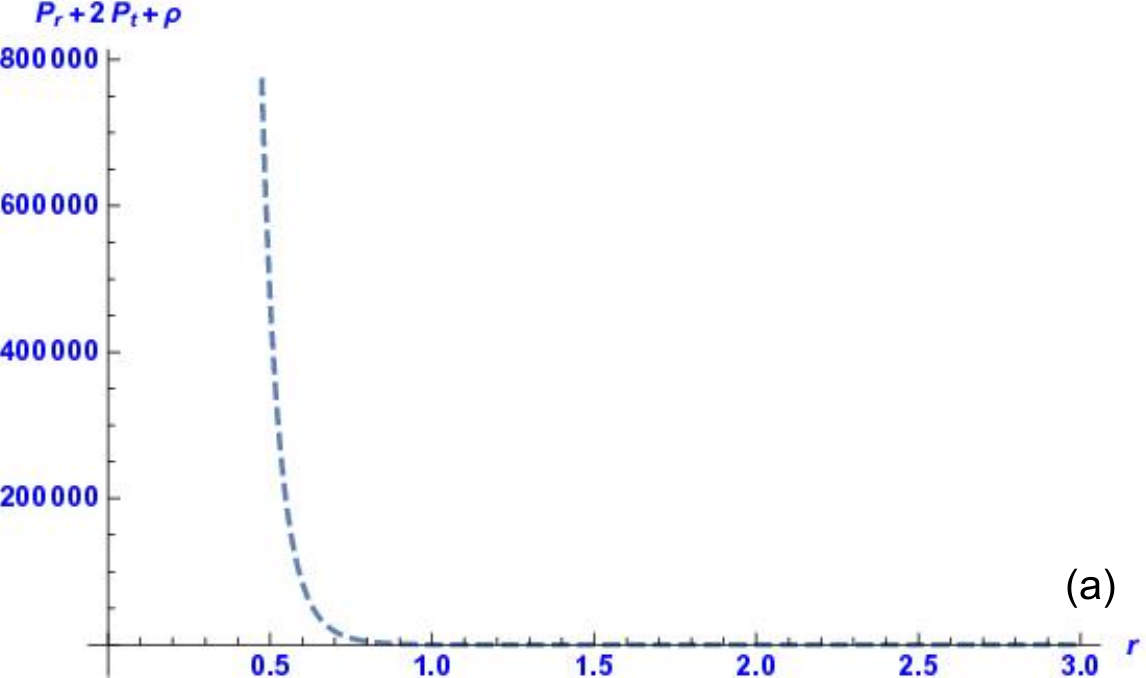}
	\vspace{1cm}
	\includegraphics[width=.40\textwidth,keepaspectratio]{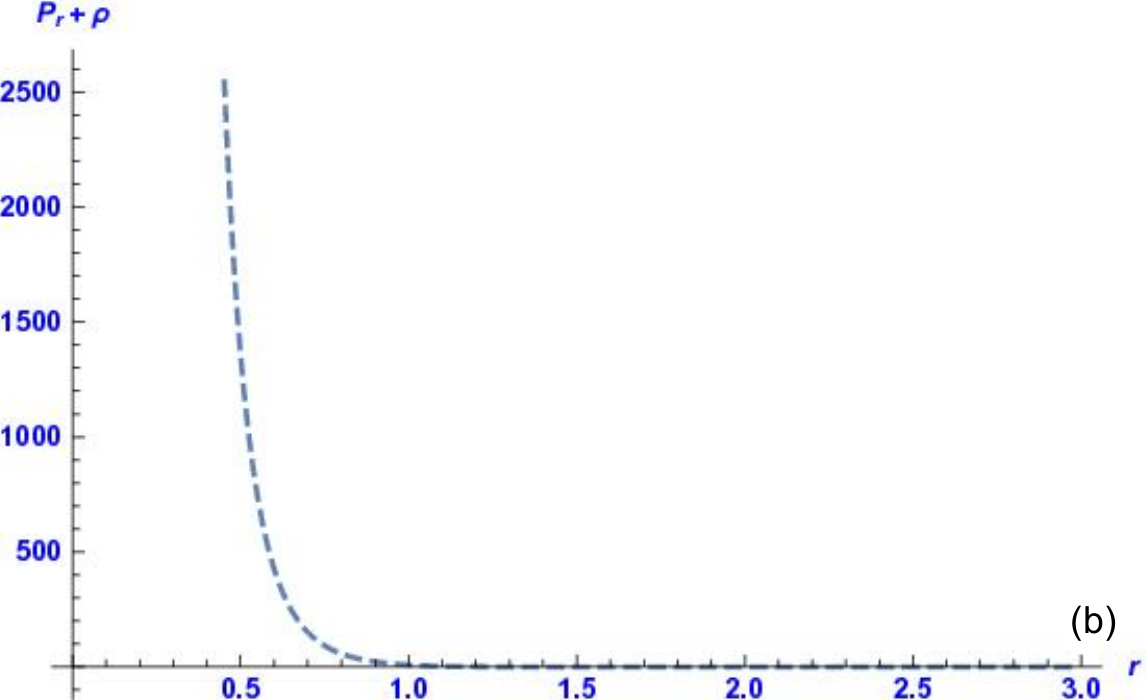}
	\hspace{1cm}
	\includegraphics[width=.40\textwidth,keepaspectratio]{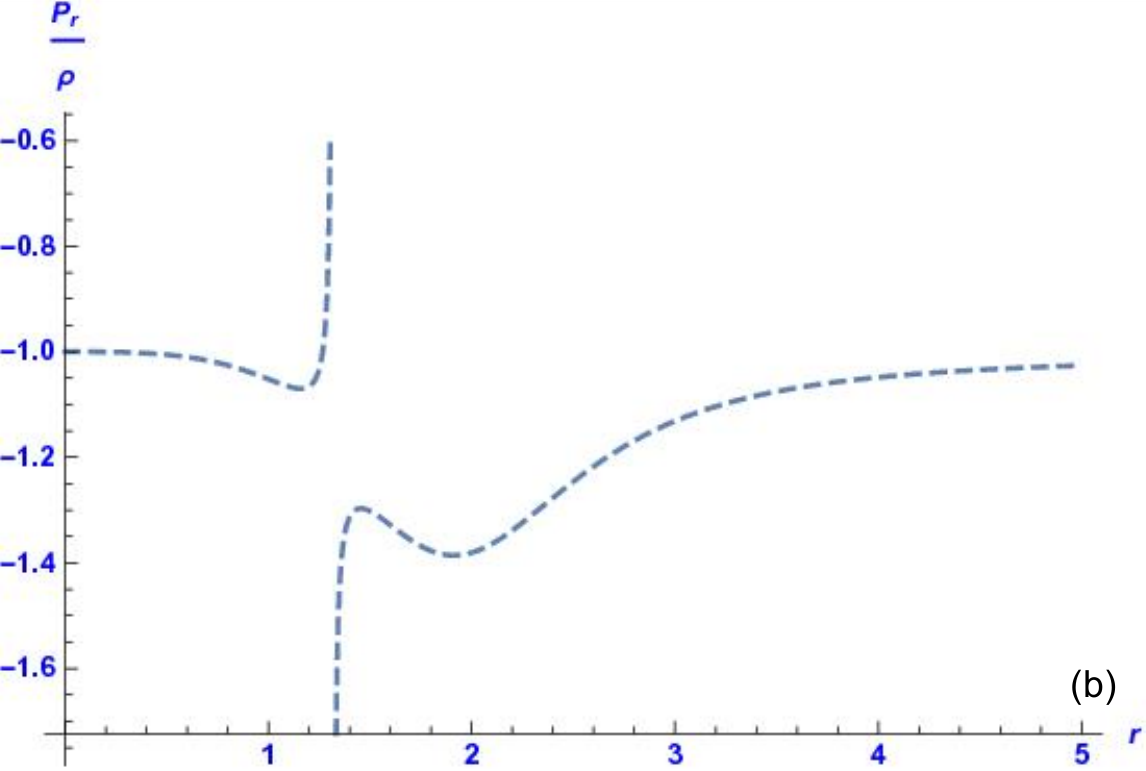}
	\includegraphics[width=.40\textwidth,keepaspectratio]{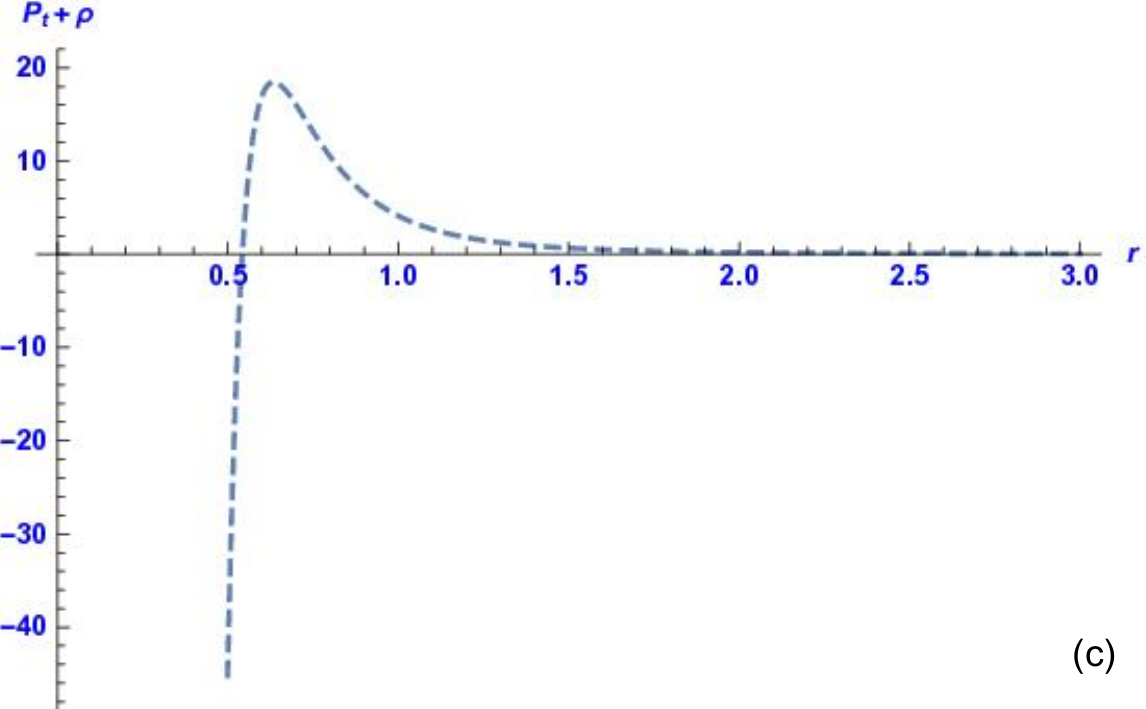}
	\hspace{1cm}
	\includegraphics[width=.40\textwidth,keepaspectratio]{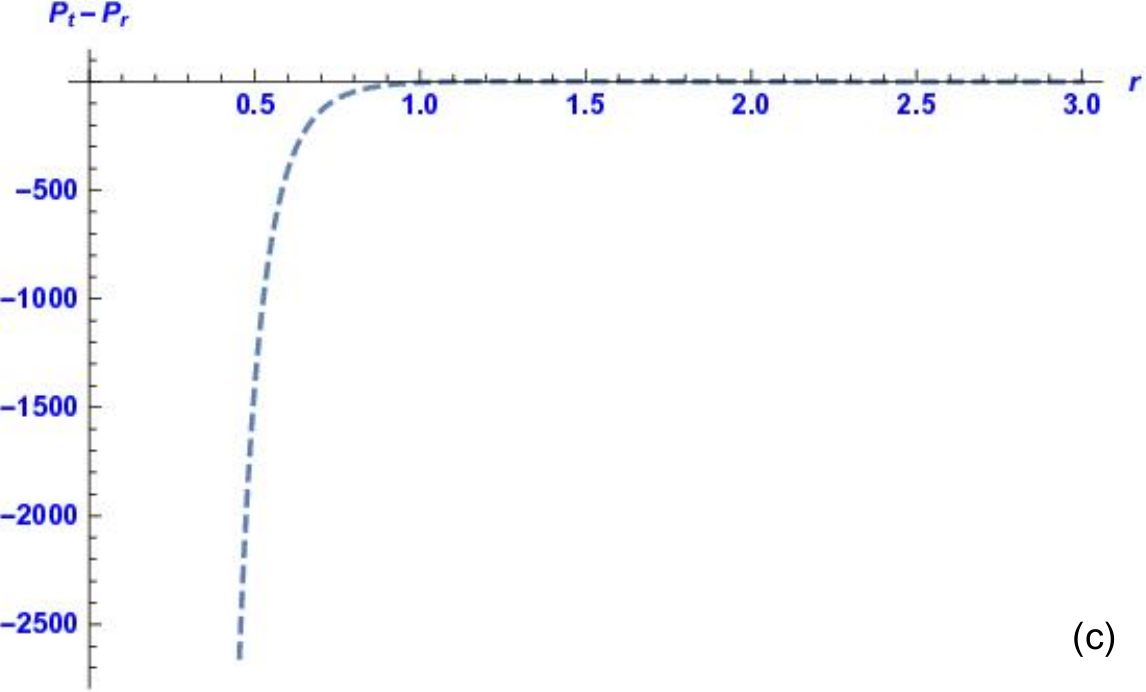}
	\includegraphics[width=.40\textwidth,keepaspectratio]{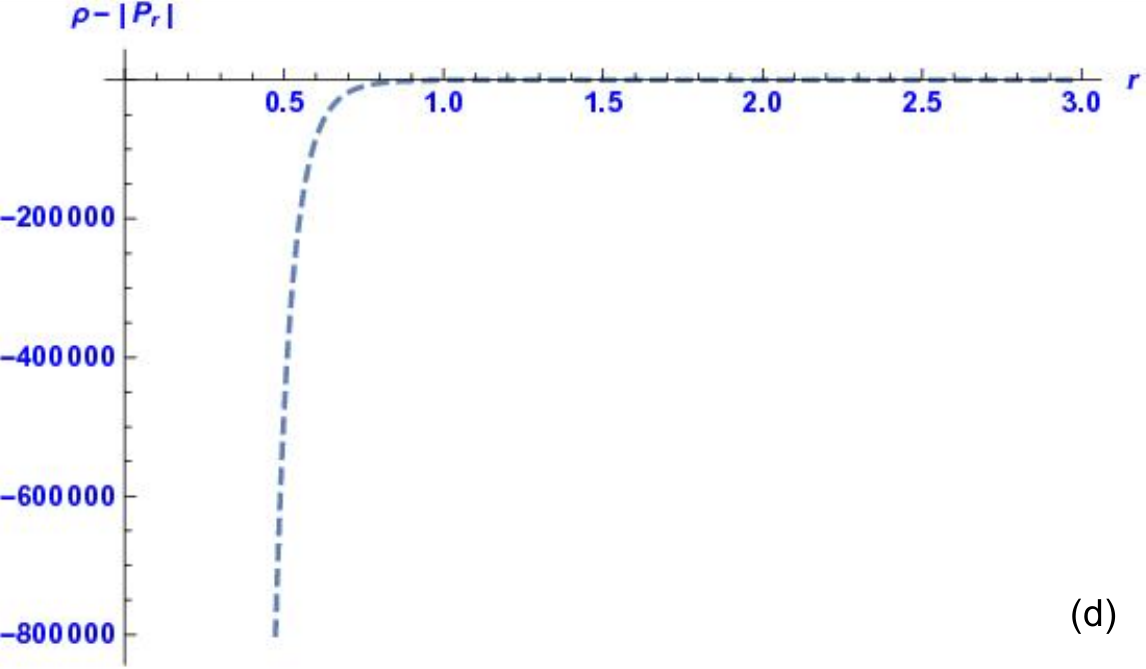}
	\hspace{1cm}
	\includegraphics[width=.40\textwidth,keepaspectratio]{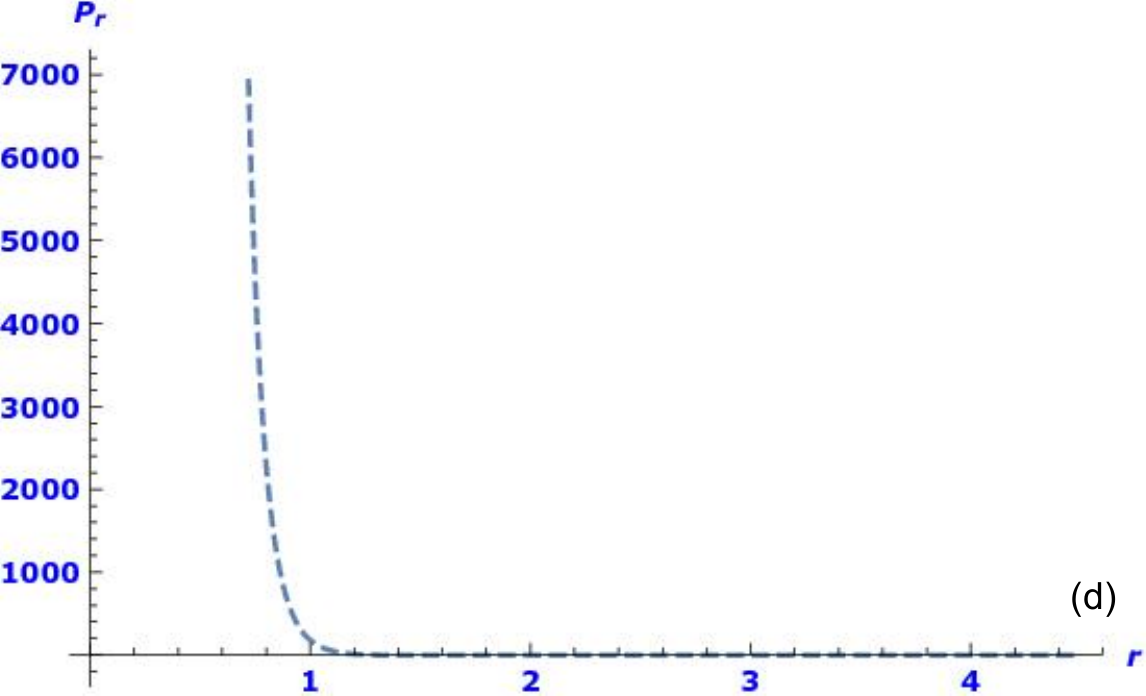}
	\includegraphics[width=.4\textwidth,keepaspectratio]{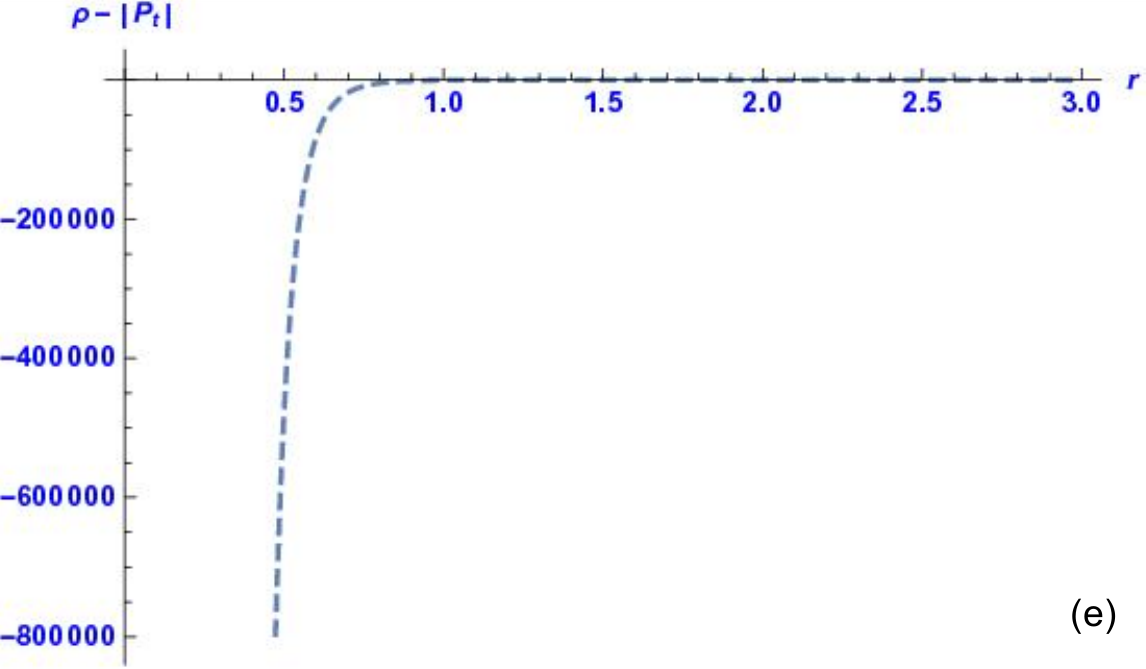}
	\hspace{0.75cm}
	\includegraphics[width=.40\textwidth,keepaspectratio]{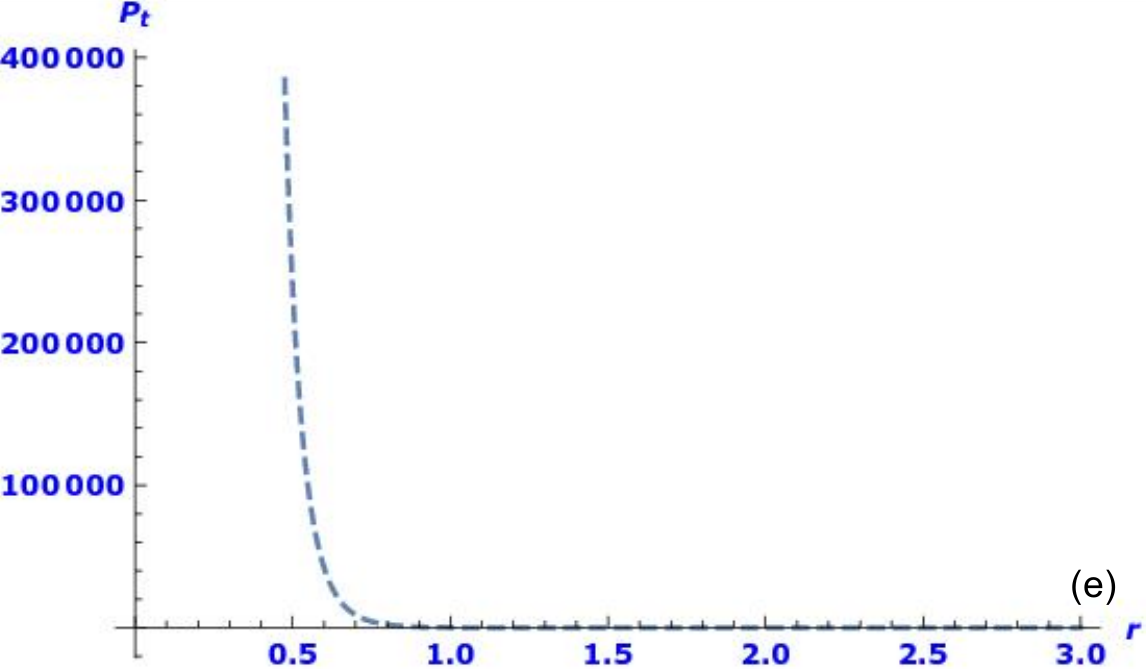}
    \caption{\textit{Left}: $\rho$, $\rho+P_{r}$, $\rho+P_{t}$, $\rho-|P_{r}|$ and $\rho-|P_{t}|$ plotted versus $r$ in the case iii. \textit{Right}: $\rho+P_{r}+2P_{t}$, $\frac{P_{r}}{\rho}$, $P_{t}-P_{r}$, $P_{r}$ and $P_{t}$ plotted versus $r$ in the case iii.}
	\label{fig3}
\end{figure*}

In the third case, the shape and redshift functions take the following form
\begin{equation}
b(r)=\frac{r_{0}\log(1+r)}{\log(1+r_{0})},\hspace{1cm}\Phi(r)=\frac{1}{r},
\label{15}
\end{equation} 
and then the parameters $\rho$, $P_{r}$ and $P_{t}$ of the model are driven as shown in (\ref{a13} - \ref{a15}). Figure \ref{fig3} shows the satisfactory situation of different energy conditions for the logarithmic corrected model in case iii (\ref{15}). Similar to the previous case, the left-side panel (a) tells us that the WEC is violated for all values of $r$ because of negative value of $\rho$. From the left-side panels (b) and (c), we find that the first condition of the NEC is satisfied while the second condition is partially fulfilled for $r>0.5$. Obviously, the left-side panels (d) and (e) reveal that the DEC is violated since $\rho-|P_{r}|$ and $\rho-|P_{t}|$ are negative. From the right-side panel (a), we find that the first and third conditions of the SEC are satisfied but the the second condition is partially satisfied just for $r>0.5$. Regarding two parameters $w$ and $\Delta$, the right-side panel (b) and (c) show a phantom-like behavior ($w<0$) for the possible exotic matter and also an attractive geometry ($\Delta<0$) for the considered wormhole. The right-side panels (d) and (e) are drawn for $P_{r}$ and $P_{t}$ versus $r$, respectively.
\subsection{Case iv}
\begin{figure*}[!hbtp]
    \centering
	\includegraphics[width=.40\textwidth,keepaspectratio]{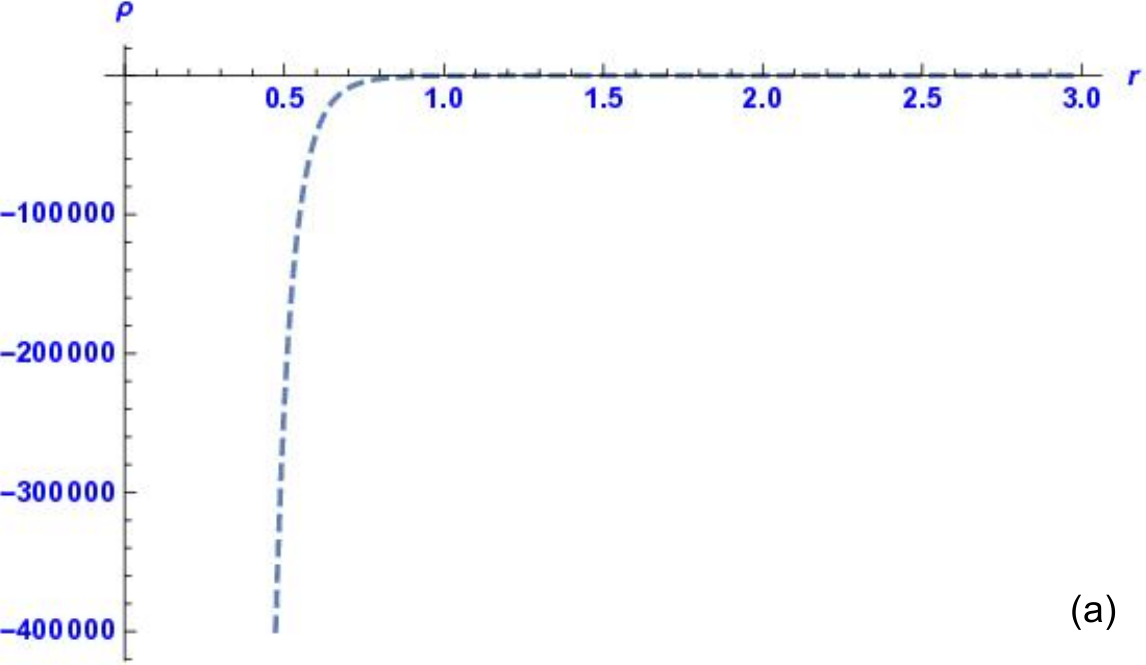}
	\hspace{1cm}
	\includegraphics[width=.40\textwidth,keepaspectratio]{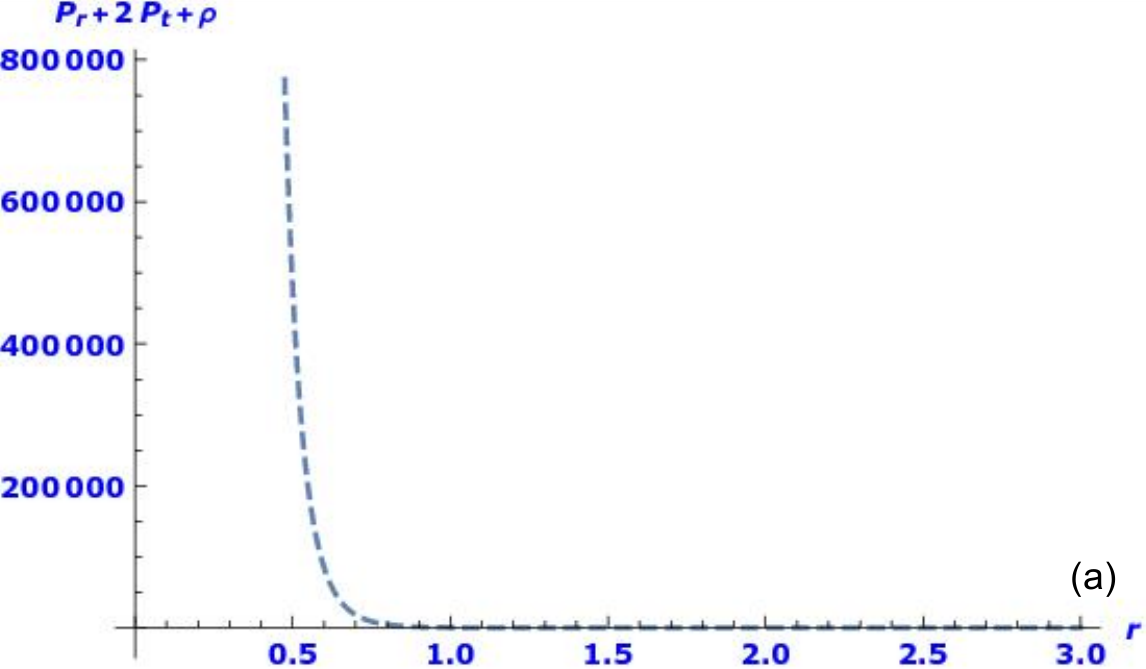}
	\vspace{1cm}
	\includegraphics[width=.40\textwidth,keepaspectratio]{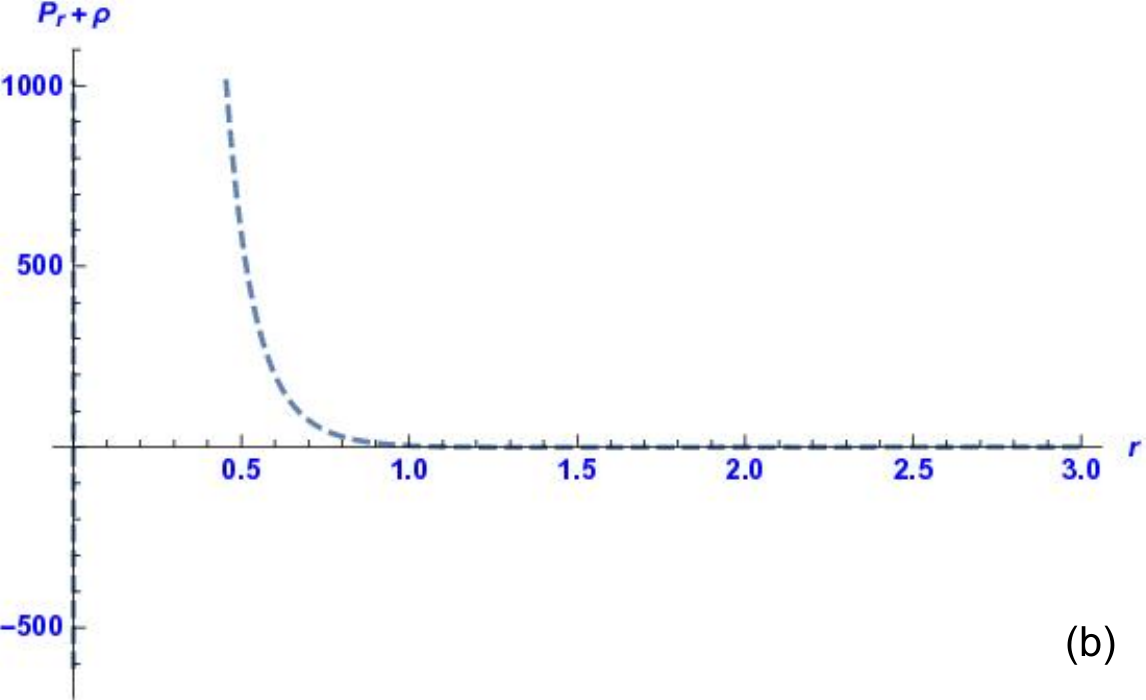}
	\hspace{1cm}
	\includegraphics[width=.40\textwidth,keepaspectratio]{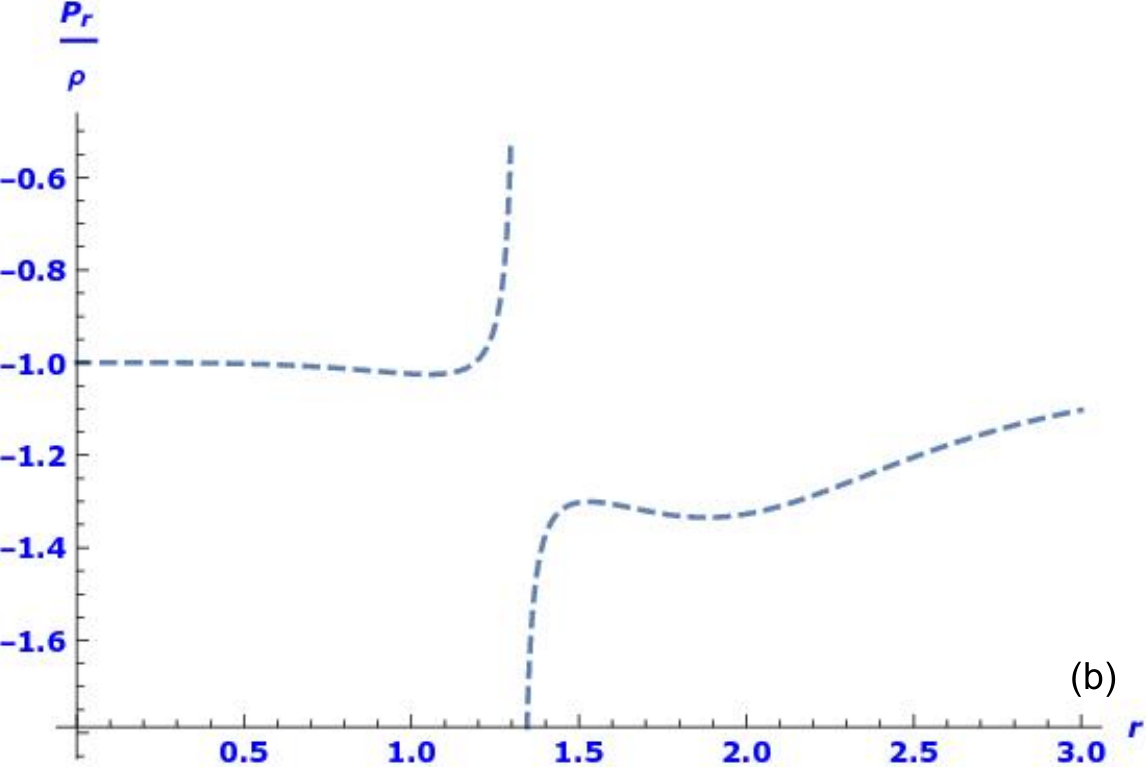}
	\includegraphics[width=.40\textwidth,keepaspectratio]{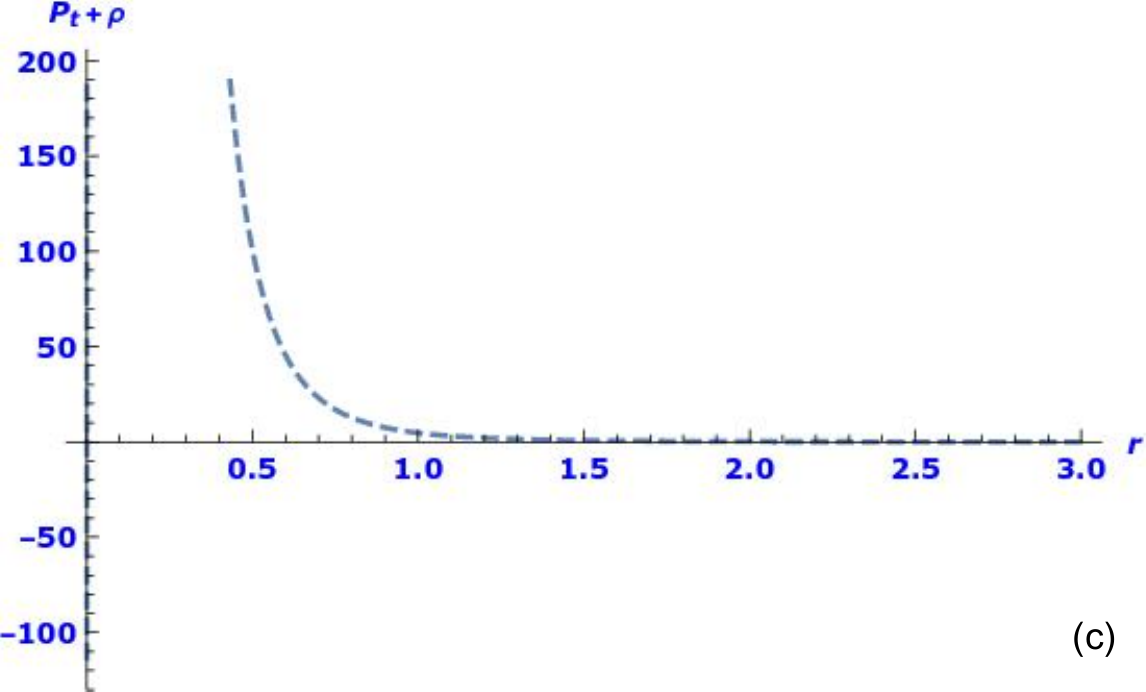}
	\hspace{1cm}
	\includegraphics[width=.40\textwidth,keepaspectratio]{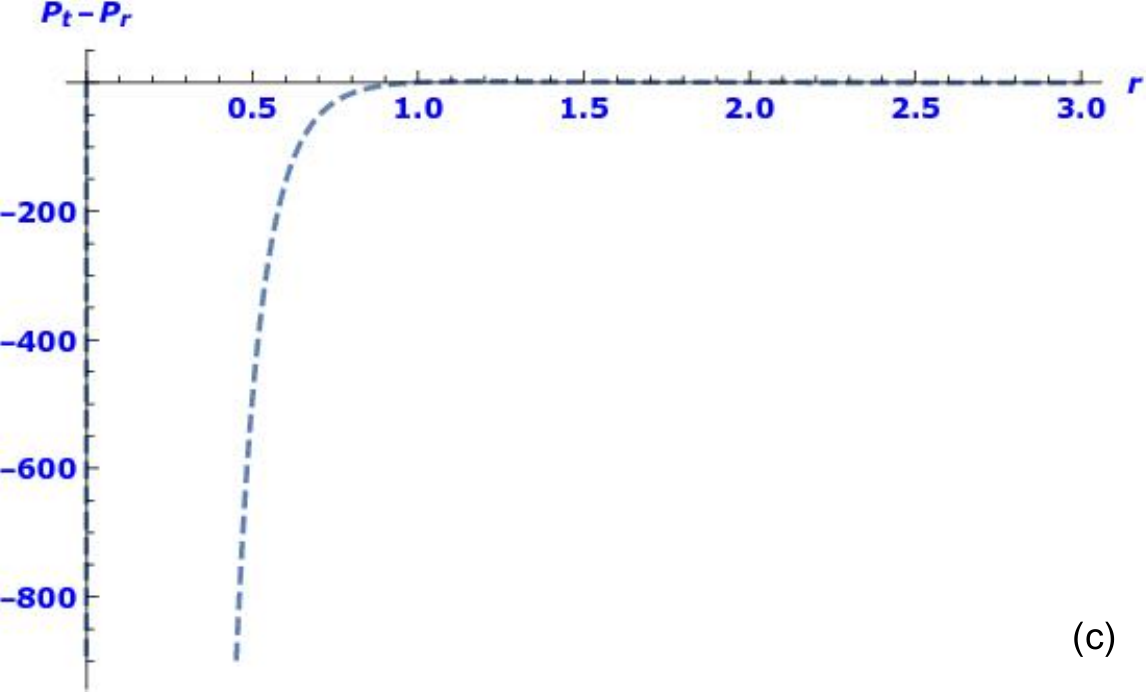}
	\includegraphics[width=.40\textwidth,keepaspectratio]{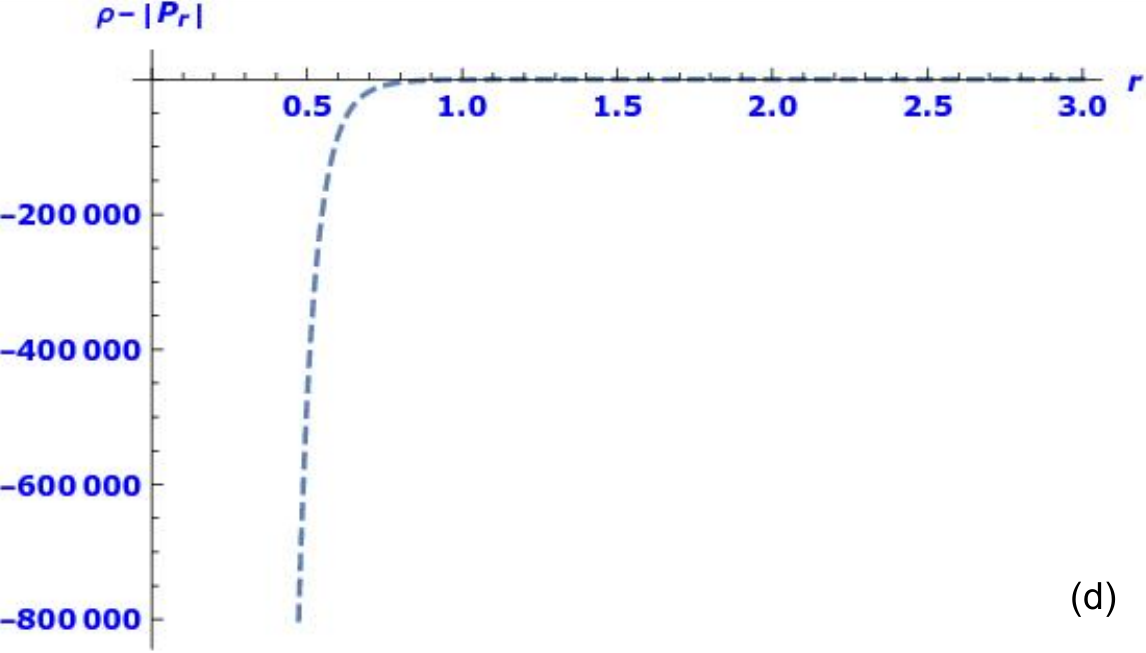}
	\hspace{1cm}
	\includegraphics[width=.40\textwidth,keepaspectratio]{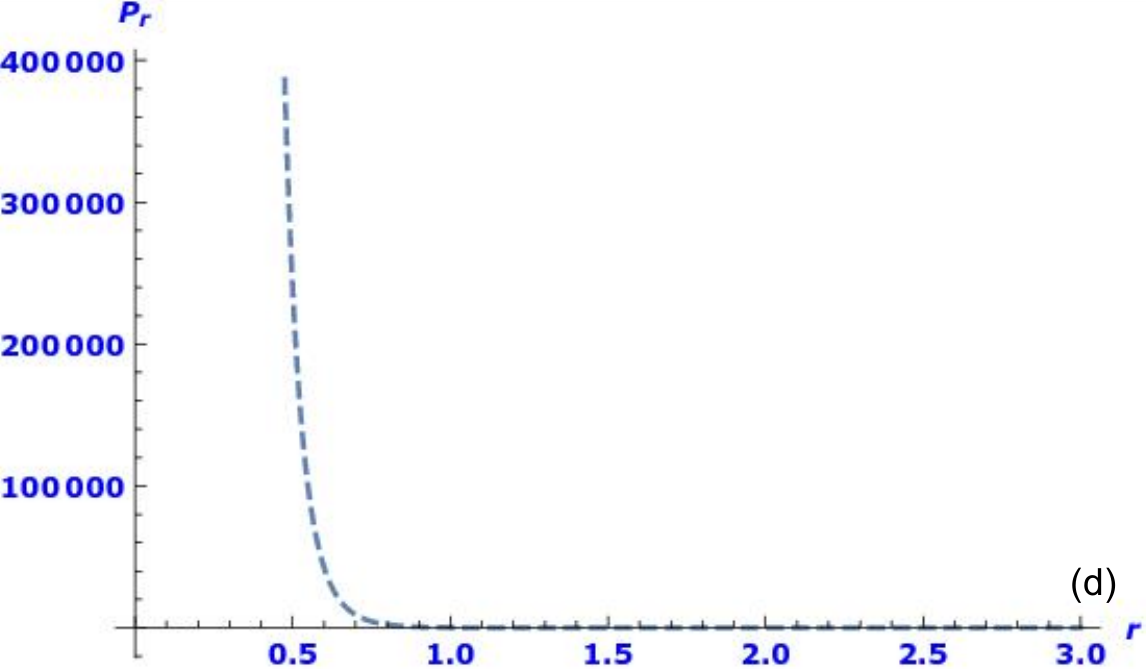}
	\includegraphics[width=.4\textwidth,keepaspectratio]{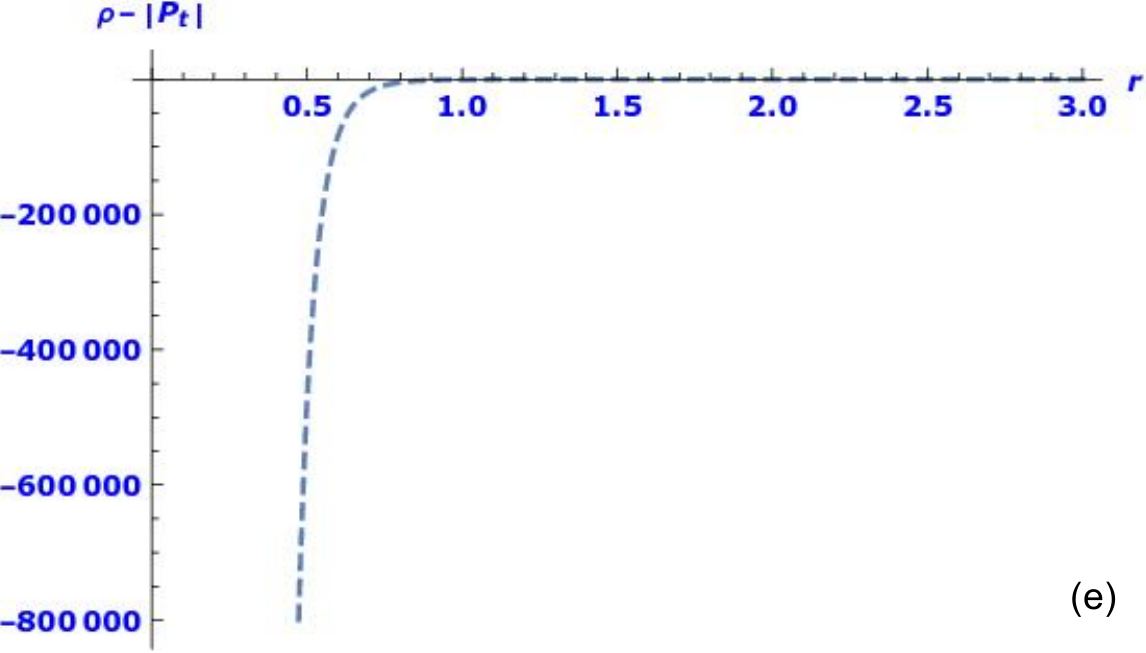}
	\hspace{0.75cm}
	\includegraphics[width=.43\textwidth,keepaspectratio]{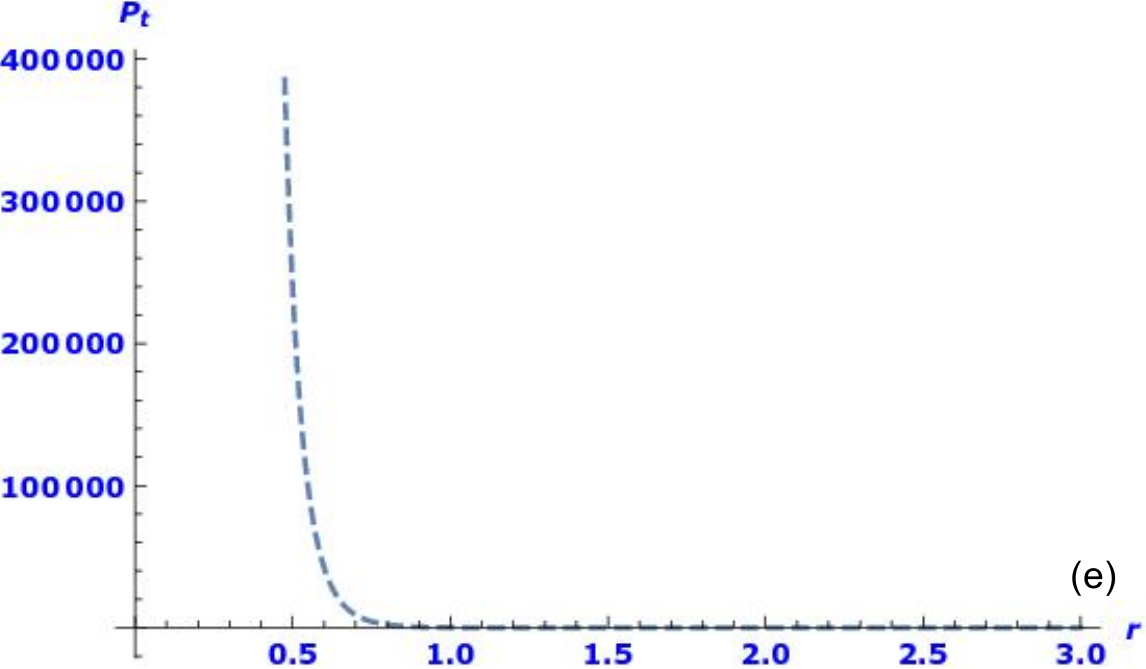}
    \caption{\textit{Left}: $\rho$, $\rho+P_{r}$, $\rho+P_{t}$, $\rho-|P_{r}|$ and $\rho-|P_{t}|$ plotted versus $r$ in the case iv. \textit{Right}: $\rho+P_{r}+2P_{t}$, $\frac{P_{r}}{\rho}$, $P_{t}-P_{r}$, $P_{r}$ and $P_{t}$ plotted versus $r$ in the case iv.}
	\label{fig4}
\end{figure*}

For the last case, we consider the shape and redshift functions as
\begin{equation}
b(r)=\frac{r_{0}\log(1+r)}{\log(1+r_{0})},\hspace{1cm}\Phi(r)=c.
\label{16}
\end{equation}
Then, the parameters $\rho$, $P_{r}$ and $P_{t}$ are calculated as shown in (\ref{a16} - \ref{a18}). In Figure \ref{fig4}, we investigate the satisfactory situation of different energy conditions for the logarithmic corrected model in case iv (\ref{16}). From the left-side panel (a), we find that WEC is violated since $\rho$ takes a negative value. The left-side panels (b) and (c) reveal that the NEC is satisfied for the model. The DEC is violated because two conditions $\rho-|P_{r}|>0$ and $\rho-|P_{r}|>0$ are invalid (the left-side panels (d) and (e)). The SEC is satisfied since all conditions are fulfilled. Moreover, the situation of two parameters $w$ and $\Delta$ in this case is similar to the previous cases in which we deal with a phantom-like ($w<0$) exotic matter in an attractive geometry ($\Delta<0$) for the wormhole (the right-side panels (b) and (c)). 
\section{conclusions}
Wormholes are known as hypothetical geometric objects that can solve the Einstein equation by violation of some important energy conditions in particular the NEC. In fact, the wormholes are able to create a geometry that connects two pints of the same space-time in order to connect two different universes. As an important key, the existence of wormholes requires an exotic matter that violates the energy conditions. This can be problematic when we face a lack of matter to tolerate the wormhole geometry. In order to remove the problem, modified theories of gravity are proposed since they deal with some modifications in geometry instead of matter. One of the most important class of such theories is called $f(R)$ gravity dealing with a general function of the Ricci scalar $R$ instead of $R$ in the Einstein-Hilbert action. In this paper, we have studied the wormholes in the context of $f(R)$ gravity by considering a logarithmic corrected $f(R)$ model for different forms of shape $b(r)$ and redshift $\Phi(r)$ functions. We have calculated the necessary parameters \textit{i.e.} energy density $\rho$, radial pressure $P_{r}$ and tangential pressure $P_{t}$ for all cases, separately. Then, we have investigated the different energy conditions including the Null Energy Condition (NEC), the Weak Energy Condition (WEC), Strong Energy Condition (SEC) and Dominant Energy Conditions (DEC) by the corresponding plots. We have carried out the analysis first for an exponential form of shape function $b(r)=\frac{r}{\exp(r-r_{0})}$ with redshift function $\Phi(r)=1/r$ (case i) and $\Phi(r)=c$ (case ii), separately. As the second case, we have focused on a logarithmic form of shape function $b(r)=\frac{r_{0}\log(1+r)}{\log(1+r_{0})}$ with redshift function $\Phi(r)=1/r$ (case iii) and $\Phi(r)=c$ (case iv). Our results can be summarized as follows:
\begin{itemize}
\item \textbf{Case i and ii}
   \begin{itemize}
      \item \textit{NEC is violated since $\rho+P_{r}\geq0$ and $\rho+P_{t}\leq0$.}
      \item \textit{WEC is violated since $\rho\leq0$, $\rho+P_{r}\geq0$ and $\rho+P_{t}\leq0$.}
      \item \textit{SEC is violated since $\rho+P_{r}\geq0$, $\rho+P_{t}\leq0$ and $\rho+P_{r}+2P_{t}\geq0$.}
      \item \textit{DEC is violated since $\rho\leq0$, $\rho-|P_{r}|<0$, $\rho-|P_{t}|<0$ and $\rho+P_{r}+2P_{t}\geq0$.} 
   \end{itemize}
\item \textbf{Case iii}
   \begin{itemize}
   \item \textit{NEC is violated since $\rho+P_{r}\geq0$ and $\rho+P_{t}\geq0$ (only for $r>0.5$).}
  \item \textit{WEC is violated since $\rho\leq0$, $\rho+P_{r}\geq0$ and $\rho+P_{t}\geq0$ (only for $r>0.5$).}
  \item \textit{SEC is violated since $\rho+P_{r}\geq0$ , $\rho+P_{t}\geq0$ (only for $r>0.5$) and $\rho+P_{r}+2P_{t}\geq0$.}
  \item \textit{DEC is violated since $\rho<0$, $\rho-|P_{r}|<0$, $\rho-|P_{t}|<0$ and $\rho+P_{r}+2P_{t}\geq0$.}
   \end{itemize} 
\item \textbf{Case iv}
   \begin{itemize}
   \item \textit{NEC is satisfied since $\rho+P_{r}\geq0$ and $\rho+P_{t}\geq0$.}
  \item \textit{WEC is violated since $\rho\leq0$, $\rho+P_{r}\geq0$ and $\rho+P_{t}\geq0$.}
  \item \textit{SEC is satisfied since $\rho+P_{r}\geq0$ , $\rho+P_{t}\geq0$ and $\rho+P_{r}+2P_{t}\geq0$.}
  \item \textit{DEC is violated since $\rho\leq0$, $\rho-|P_{r}|<0$, $\rho-|P_{t}|<0$ and $\rho+P_{r}+2P_{t}\geq0$.}
   \end{itemize}
\end{itemize}
As a piece of additional information, we have found that all cases show a negative value of $w$ and $\Delta$ that the first implies the existence of an exotic matter with a phantom-like behavior and the second refers to an attractive geometry for the wormhole. In this paper, we have shown that the wormholes introduced in the context of a logarithmic-corrected model of $f(R)$ gravity with the corresponding shape and redshift functions (except in case iv for the NEC and SEC) can violate all energy conditions due to the existence of an exotic matter generated from the modifications in an attractive geometry. 
\bibliographystyle{ieeetr}
\bibliography{biblo}
\newpage
\appendix
\section{Energy density and pressure parameters}
\subsection{Case i}
\begin{eqnarray}
&\!&\!\rho=\frac{1}{2r^{7}}e^{-3r}(\frac{2^{1+2n}e^{4r-2r_{0}}(e^{r}-e^{r_{0}})(-1+n)^{2}n^{2}(-1+2n)r^{8}(\frac{-e^{-r+r_{0}}(-1+r)}{r^{2}})^{2n}}{(-1+r)^{2}}+2^{n}e^{r}(-\frac{e^{-r+r_{0}}(-1+r)}{r^{2}})^{n-1}r^{2}\times\nonumber\\&\!&\!
\times(16e^{2r_{0}}(n-1)n(7+5n)(-1+r)+e^{2r}(-1+n)n(-2+r)r^{2}+e^{r+r_{0}}(16(-1+n)n(7+5n)-16(-1+n)n(7+5n)\times\nonumber\\&\!&\!
\times r+2(-1+n)nr^{2}+r^{3}+(-1+(-1+n)n)r^{4}))+2e^{r_{0}}(-1+r)(-1520e^{2r_{0}}(-1+r)-e^{2r}r^{2}(-20+10r+r^{3}-2e^{r+r_{0}}\times\nonumber\\&\!&\!
\times(760+r(-760+r(10+r+4r^{2}))))+4(2^{3+n}e^{2r}(e^{r}-e^{r_{0}})n(-1+n^{2})(-\frac{e^{-r+r_{0}}(-1+r)}{r^{2}})^{n}r^{4}+e^{r}(-1+r)(-544e^{2r_{0}}\times\nonumber\\&\!&\!
\times(-1+r)-2e^{2r}(-2+r)r^{2}-e^{r+r_{0}}(544+r(4+r+r^{2})))))\log(-\frac{2e^{-r+r_{0}}(-1+r)}{r^{2}})+348e^{2r_{0}}(e^{r}-e^{r_{0}})(-1+r)^{2}\times\nonumber\\&\!&\!
\times\log(-\frac{2(-1+r)e^{-r+r_{0}}}{r^{2}})^{2}),
\label{a1}
\end{eqnarray}
\begin{eqnarray}
&\!&\!P_{r}=\frac{1}{2r^{7}}(3040e^{-3r+3r_{0}}(-1+r)^{2}-\frac{2^{1+2n}e^{2(r-r_{0})}(-1+n)^{2}n^{2}(-1+2n)r^{8}(-\frac{e^{-r+r_{0}}(-1+r)}{r^{2}})^{2n}}{(-1+r)^{2}}-2e^{-r+r_{0}}r^{2}(-12+r\times\nonumber\\&\!&\!
\times
(-74+(88+2^{3+n}(-1+n)n(7+5n)(-\frac{e^{-r+r_{0}}(-1+r)}{r^{2}})^{n}-r)r))4e^{-2r+2r_{0}}(-1+r)(760+r(-760+r(-6+r\times\nonumber\\&\!&\!
\times
(-37+12r))))+4e^{-3r}\log(-\frac{2e^{-r+r_{0}}(-1+r)}{r^{2}})(2^{3+n}e^{2r}(-e^{r}+e^{r_{0}})n(-1+n^{2})(-\frac{e^{-r+r_{0}}(-1+r)}{r^{2}})^{n}r^{4}e^{r_{0}}(-1+r)\times\nonumber\\&\!&\!
\times
(544e^{2r_{0}}(-1+r)+2e^{2r}r^{2}(2+7r)+e^{r+r_{0}}(544+r(-544+r(-4+r(-11+5r))))+96e^{r_{0}}(+e^{r_{0}}-e^{r})(-1+r)\times\nonumber\\&\!&\!
\times\log(-\frac{2(-1+r)e^{-r+r_{0}}}{r^{2}}))))+\frac{A+B}{-1+r},
\label{a2}
\end{eqnarray}
where
\begin{equation}
A=r^{4}(4-4r-2^{n}(-\frac{e^{-r+r_{0}}(-1+r)}{r^{2}})^{n}(-16(-1+n)n(7+5n)
\label{a3}
\end{equation}
and
\begin{equation}
B=16(-1+n)n(7+5n)r+2nr^{2}+(-1+2(-1+n)n(-1+2n))r^{3}+(1+n-2n^{2})r^{4})),
\label{a4}
\end{equation}
\begin{eqnarray}
&\!&\!P_{t}=\frac{1}{4r^{7}}e^{-3r}(6080e^{3r_{0}}(-1+r)^{2}-\frac{4^{n+1}e^{5r-2r_{0}}(-1+n)^{2}n^{2}(-1+2n)r^{8}(\frac{-e^{-r+r_{0}}(-1+r)}{r^{2}})^{2n}}{(-1+r)^{2}}+\nonumber\\&\!&\!
+\frac{2^{n+1}e^{4r-r_{0}}nr^{5}(-\frac{e^{-r+r_{0}}(-1+r)}{r^{2}})^{n}(1+r^{2}(-2+n+r+n(-1+2^{1+n}(-1+n)^{2}(-1+2n)(-\frac{e^{-r+r_{0}}(-1+r}{r^{2}})^{n})r))}{(-1+r)^{2}}+\nonumber\\&\!&\!
+2e^{2r+r_{0}}r(12+r^{2}(6+r(-22+2^{4+n}(-1+n)n(7+5n)(-\frac{e^{-r+r_{0}}(-1+r}{r^{2}})^{n}+(-1+r)r)))+4e^{r+2r_{0}}(-1+r)(1520+\nonumber\\&\!&\!
+r(-1514+r(6+r(19+11r))))+16\log(-\frac{2e^{-r+r_{0}}(-1+r}{r^{2}})(-2^{2+n}e^{2r}(e^{r}-e^{r_{0}})n(-1+n^{2})(-\frac{e^{-r+r_{0}}(-1+r}{r^{2}})^{n}r^{4}+\nonumber\\&\!&\!
+e^{r_{0}}(-1+r)(272e^{2r_{0}}(-1+r)-e^{2r}r(1+r+r^{2})+e^{r+r_{0}}(272+r(-271+r(1+r)^{2}))+48e^{r_{0}}(-e^{r}-e^{r_{0}})(-1+r)\times\nonumber\\&\!&\!
\times
\log(-\frac{2e^{-r+r_{0}}(-1+r}{r^{2}}))))+\frac{A+B}{-1+r}
\label{a5}
\end{eqnarray}
where
\begin{equation}
A=e^{3r}r^{3}(-4+r(4r+2^{n}(-\frac{e^{-r+r_{0}}(-1+r}{r^{2}})^n(32(-1+n)n(7+5n)-2n(-113+16n(2+5n))r
\label{a6}
\end{equation}
and
\begin{equation}
B=2nr^{2}+(2+n(-5+4n))r^{3}+(-2+n(-1+2n))r^{4})))
\label{a7}
\end{equation}
\subsection{Case ii}
\begin{eqnarray}
&\!&\!\rho=\frac{1}{2}(2^{2}(-\frac{2e^{-r+r_{0}}(-1+r)}{r^{2}})^{n}-\frac{20e^{-r+r_{0}}(-1+r)}{r^{4}}+\frac{4e^{-2r+2r_{0}}(-1+r)^{2}}{r^{4}}-\frac{2e^{-r+r_{0}}(-1+r)}{r^{2}}+\nonumber\\&\!&\!
+\frac{2^{n}(-1+n)n(-\frac{2e^{-r+r_{0}}(-1+r)}{r^{2}})^{-1+n}}{r^{2}}-\frac{8e^{-r+r_{0}}(-1+r)\log(-\frac{e^{2-r+r_{0}}(-1+r)}{r^{2}})}{r^{4}}+\nonumber\\&\!&\!
+\frac{4e^{-2r+2r_{0}}(-1+r)^{2}\log(-\frac{2e^{-r+r_{0}}(-1+r)}{r^{2}})}{r^{4}}+\frac{A*B}{r^{3}},
\label{a8}
\end{eqnarray}
where
\begin{equation}
A=2(1-e^{-r+r_{0}}(-20e^{-r+r_{0}}(-1+r)+2^{n}(-1+n)n(-\frac{e^{-r+r_{0}}(-1+r)}{r^{2}})^{n-1}r^{2}8e^{-r+r_{0}}(-1+r)\log(-\frac{2e^{-r+r_{0}}(-1+r)}{r^{2}})
\label{a9}
\end{equation}
and
\begin{equation}
B=(\frac{1}{-2+2e^{r-r_{0}}}+\frac{2^{n}(-1+n)n(-1+2n)(-\frac{e^{-r+r_{0}}(-1+r)}{r^{2}})^{n-1}}{r^{2}}-\frac{4e^{-r+r_{0}}(-1+r)(19+6\log(-\frac{2e^{-r+r_{0}}(-1+r)}{r^{2}})}{r^{4}}
\label{a10}
\end{equation}
\begin{eqnarray}
&\!&\!P_{r}=\frac{1}{2r^{7}}e^{-3r}(3040e^{3r_{0}}(-1+r)^{2}+\frac{2^{2n+1}e^{4r-2r_{0}}(-e^{r}+e^{r_{0}})(-1+n)^{2}n^{2}(-1+2n)r^{8}(-\frac{e^{-r+r_{0}}(-1+r)}{r^{2}})^{2n}}{(-1+r)^{2}}-2e^{2r+r_{0}}\times\nonumber\\&\!&\!
\times(86+(-86+r))+4e^{r+2r_{0}}(-1+r)(760-760r-37r^{3}+12r^{4})+2^{n}e^{r_{0}}(-1+r)(-\frac{e^{-r+r_{0}}(-1+r)}{r^{2}})^{-2+n}(16e^{2r_{0}}\times\nonumber\\&\!&\!
\times(-1+n)n(7+5n)(-1+r)+e^{2r}(-1+n)n(-1+4n)r^{3}e^{r+r_{0}}(16(-1+n)n(7+5n)-16(-1+n)n(7+5n)r+(1-\nonumber\\&\!&\!
-2(-1+n)n(-1+2n))r^{3}(-1+n)(1+2n)r^{4}))+4(2^{3+n}e^{2r}(-e^{r}+e^{r_{0}})n(-1+n^{2})(-\frac{e^{-r+r_{0}}(-1+r)}{r^{2}})^{n}r^{4}+e^{r_{0}}\times\nonumber\\&\!&\!
\times(-1+r)(544e^{2r_{0}}(-1+r)+14e^{2r}r^{3}+e^{r+r_{0}}(544-544r-11r^{3}+5r^{4})))\log(-\frac{2e^{-r+r_{0}}(-1+r)}{r^{2}})+384e^{2R_{0}}\times\nonumber\\&\!&\!
\times(-e^{r}+e^{r_{0}})(-1+r^{2})\log(-\frac{e^{-r+r_{0}}(-1+r)}{r^{2}})^{2})
\label{a11}
\end{eqnarray}
\begin{eqnarray}
&\!&\!P_{t}=\frac{1}{4r^{7}}e^{-3r}(\frac{4^{1+n}e^{4r-2r_{0}}(-e^{r}+e^{r_{0}}(-1+n)^{2}n^{2}(-1+2n)r^{8}(-\frac{e^{-r+r_{0}}(-1+r)}{r^{2}})^{2n}}{(-1+r)^{2}}+2e^{r_{0}}(3040e^{2r_{0}}(-1+r)^{2}+\nonumber\\&\!&\!
+e^{2r}r^{3}(20-20r+r^{3})+2e^{r+r_{0}}(-1+r)(1520-1520r+16r^{3}+11r^{4}))+2^{n}e^{r_{0}}(-1+r)(-\frac{e^{-r+r_{0}}(-1+r)}{r^{2}})^{-2+n}\times\nonumber\\&\!&\!
\times(32e^{2r_{0}}(-1+n)n(7+5n)(-1+r)-2e^{r_{0}}(-1+n)nr^{3}+e^{r+r_{0}}(32(-1+n)n(7+5n)-32(-1+n)n(7+5n)r+2\times\nonumber\\&\!&\!
\times(-1+n)(-1+2n)r^{3}+(-2+n(-1+2n))r^{4}))+8(2^{3+n}e^{2r}(-e^{r}+e^{r_{0}})n(-1+n^{2})(-\frac{e^{-r+r_{0}}(-1+r)}{r^{2}})^{n}r^{4}+e^{r_{0}}\times\nonumber\\&\!&\!
\times(-1+r)(544+e^{2r_{0}}(-1+r)(544-544r+3r^{3}+2r^{4})))\log(-\frac{2e^{-r+r_{0}})(-1+r)}{r^{2}})+768e^{2r_{0}}(+e^{r_{0}}-e^{r})(-1+r)^{2}\times\nonumber\\&\!&\!
\times\log(-\frac{2e^{-r+r_{0}})(-1+r)}{r^{2}})^{2})
\label{a12}
\end{eqnarray}
\subsection{Case iii}
\begin{eqnarray}
&\!&\!\rho=\frac{1}{2r^{8}(1+r)^{2}r_{0}^{2}\log(1+r_{0})^{3}}(2^{1+2n}(-1+n)^{2}n^{2}(-1+2n)r^{9}(1+r)^{4}(\frac{r_{0}}{r^{2}(1+r)\log(1+r_{0})})^{2n}\log(1+r_{0})^{5}-r_{0}\times\nonumber\\&\!&\!
\times\log(1+r)(2^{n}(-1+n)nr^{4}(1+r)^{2}(\frac{r_{0}}{r^{2}(1+r)\log(1+r_{0})})^{n}\log(1+r_{0})^{2}+4r_{0}^{2}(5+2(\frac{r_{0}}{r^{2}(1+r)\log(1+r_{0})})))(-r^{2}\times\nonumber\\&\!&\!
\times(1+r)(2+r)r_{0}\log(1+r_{0})+2^{1+n}(-1+n)n(-1+2n)r^{4}(1+r)^{2}(\frac{r_{0}}{r^{2}(1+r)\log(1+r_{0})})^{n}\log(1+R_{0})^{2}+8r_{0}^{2}(19+\nonumber\\&\!&\!
+6\log(\frac{2r_{0}}{r^{2}(1+r)\log(1+r_{0})})))+2^{n}r^{5}(1+r)^{2}r_{0}(\frac{r_{0}}{r^{2}(1+r)\log(1+r_{0})})^{n}\log(1+r_{0})^{3}((r^{3}+(-1+n)n(112+80n-\nonumber\\&\!&\!
-r^{3}))r_{0}+(-1+n)((-2+r)r^{2}(1+r)\log(1+r_{0})+32(1+n)r_{0}\log(\frac{2r_{0}}{r^{2}(1+r)\log(1+r_{0})})))+2rr_{0}^{3}\log(1+r_{0})(r^{2}\times\nonumber\\&\!&\!
\times(1+r)\log(1+r_{0})(-20+10r+r^{3}+4(-2+r)\log(\frac{2r_{0}}{r^{2}(1+r)\log(1+r_{0})}))+2r_{0}(760-4r^{3}+\nonumber\\&\!&\!
+\log(\frac{r_{0}}{2r^{2}(1+r)\log(1+r_{0})})(544-r^{3}+96\log(\frac{2r_{0}}{r^{2}(1+r)\log(1+r_{0})})))))
\label{a13}
\end{eqnarray}
\begin{eqnarray}
&\!&\!P_{r}=\frac{1}{2r^{8}(1+r)^{2}r_{0}\log(1+r_{0})^{3}}(-2^{1+2n}(-1+n)^{2}n^{2}(-1+2n)(1+r)^{3}r^{7}(\frac{r_{0}}{r^{2}(1+r)\log(1+r_{0})})^{-1+2n}\log(1+r_{0})^{4}-\nonumber\\&\!&\!
-2^{n}r^{5}(1+r)^{2}(\frac{r_{0}}{r^{2}(1+r)\log(1+r_{0})})^{n}\log(1+r_{0})^{3}((-1+n)(16n(7+5n)-(1+2n)r^{3})r_{0}+nr^{2}(1+r)(2+(-1+n)\times\nonumber\\&\!&\!
\times(-1+4n)r)\log(1+r_{0})+32n(-1+n^{2})r_{0}\log(\frac{2r_{0}}{r^{2}(1+r)\log(1+r_{0})}))+4rr_{0}\log(1+r_{0})(4(-190+3r^{3})r_{0}^{2}-r^{2}(1+r)\times\nonumber\\&\!&\!
\times\log(1+r_{0})((6+43r)r_{0}+r^{2}(1+r)\log(1+r_{0}))+r_{0}((-544+5r^{3})r_{0}-2r^{2}(1+r)(2+7r)\log(1+r_{0}))\times\nonumber\\&\!&\!
\times\log(\frac{2r_{0}}{r^{2}(1+r)\log(1+r_{0})})-96r_{0}^{2}\log(\frac{2r_{0}}{r^{2}(1+r)\log(1+r_{0})})^{2})+\log(1+r_{0})(2^{1+2n}(-1+n)^{2}n^{2}(-1+2n)r^{8}(1+r)^{4}\times\nonumber\\&\!&\!
\times(\frac{2r_{0}}{r^{2}(1+r)\log(1+r_{0})})^{2n}\log(1+r_{0})^{4}2r_{0}^{2}(1520r_{0}^{2}+r^{2}(1+r)\log(1+r_{0})(2(6+25r)r_{0}+r^{2}(2+r-r^{2})\log(1+r_{0}))+\nonumber\\&\!&\!
+4r_{0}(227r_{0}+r^{2}(1+r)(2+3r)\log(1+r_{0}))\log(\frac{2r_{0}}{r^{2}(1+r)\log(1+r_{0})})+192r_{0}^{2}\log(\frac{2r_{0}}{r^{2}(1+r)\log(1+r_{0})})^{2})+2^{n}nr^{4}\times\nonumber\\&\!&\!
\times(1+r)^{2}r_{0}(\frac{r_{0}}{r^{2}(1+r)\log(1+r_{0})})^{n}\log(1+r_{0})^{2}(r^{2}(1+r)(2+(3+4(-2+n)n)r)\log(1+r_{0})+16r_{0}(-7+2n+5n^{2}+\nonumber\\&\!&\!
+2(-1+n^{2})\log(\frac{2r_{0}}{r^{2}(1+r)\log(1+r_{0})})))))
\label{a14}
\end{eqnarray}
\begin{eqnarray}
&\!&\!P_{t}=-\frac{1}{2r^{8}(1+r)^{2}r_{0}\log(1+r_{0})^{3}}(2^{1+2n}(-1+n)^{2}n^{2}(-1+2n)r^{7}(1+r)^{3}(\frac{2r_{0}}{r^{2}(1+r)\log(1+r_{0})})^{-1+2n}\log(1+r_{0})^{4}rr_{0}\times\nonumber\\&\!&\!
\times\log(1+r_{0})(-2(-1520+r^{2}(3+11r)r_{0}^{2}+r(-12+r(-24+r(-32-21r+r^{3})))r_{0}\log(1+r_{0})-2(r+r^{2})^{3}\log(1+r_{0})-\nonumber\\&\!&\!
-4r_{0}((-544+r^{2}+2r^{3})r_{0}+2r(1+r)(1+r+r^{2})\log(1+r_{0}))\log(\frac{2r_{0}}{r^{2}(1+r)\log(1+r_{0})})+384r_{0}^{2}\times\nonumber\\&\!&\!
\times\log(\frac{2r_{0}}{r^{2}(1+r)\log(1+r_{0})})^{2})-2^{n-1}r^{5}(1+r)^{2}(\frac{r_{0}}{r^{2}(1+r)\log(1+r_{0})})^{n}\log(1+r_{0})^{3}(2nr(1+r)(1+r+(-1+n)r^{2})\times\nonumber\\&\!&\!
\times\log(1+r_{0})+r_{0}(-32(-1+n)n(7+5n)+nr^{2}+(-2+n(-1+2n))r^{3}-64n(-1+n^{2})\log(\frac{2r_{0}}{r^{2}(1+r)\log(1+r_{0})})))+\nonumber\\&\!&\!
+\frac{1}{2}\log(1+r)(-4^{n+1}(-1+n)^{2}n^{2}(-1+2n)r^{8}(1+r)^{4}(\frac{2r_{0}}{r^{2}(1+r)\log(1+r_{0})})^{2n}\log(1+r_{0})^{4}2r_{0}^{2}(-3040r_{0}^{2}+r(1+r)\times\nonumber\\&\!&\!
\times\log(1+r_{0})(6(2+3r+9r^{2})r_{0}-r^{2}(1+r)(-2+(-3+r)r)\log(1+r_{0}))+4r_{0}(-544r_{0})+r(1+r)(2+r(3+5r))\times\nonumber\\&\!&\!
\times\log(1+r_{0}))\log(\frac{2r_{0}}{r^{2}(1+r)\log(1+r_{0})})-384r_{0}^{2}\log(\frac{2r_{0}}{r^{2}(1+r)\log(1+r_{0})})^{2})+2^{n}nr^{4}(1+r)^{2}r_{0}(\frac{2r_{0}}{r^{2}(1+r)\log(1+r_{0})})^{n}\times\nonumber\\&\!&\!
\times\log(1+r_{0})^{2}(r(1+r)(2+r(3-7r+6nr))\log(1+r_{0})+32r_{0}(7-2n-5n^{2}-2(-1+n^{2})\log(\frac{2r_{0}}{r^{2}(1+r)\log(1+r_{0})})))))
\label{a15}
\end{eqnarray}
\subsection{Case iv}
\begin{eqnarray}
&\!&\!\rho=\frac{1}{2r^{8}(1+r)^{2}r_{0}^{2}\log(1+r_{0})^{3}}(4r^{4}r_{0}^{4}\log(1+r_{0})+2r^{6}(1+r)r_{0}^{3}\log(1+r_{0})^{3}2^{n}r^{8}(1+r)^{2}r_{0}^{2}(\frac{r_{0}}{r^{2}(1+r)\log(1+r_{0})})^{n}\times\nonumber\\&\!&\!
\times\log(1+r_{0})^{3}+2^{n}(-1+n)nr^{8}(1+r)^{3}r_{0}(\frac{r_{0}}{r^{2}(1+r)\log(1+r_{0})})^{n}\log(1+r_{0})^{4}+4r^{4}r_{0}^{4}\log(1+r_{0})\times\nonumber\\&\!&\!
\times\log(\frac{2r_{0}}{r^{2}(1+r)\log(1+r_{0})})+4r^{4}(1+r)r_{0}^{3}\log(1+r_{0})^{2}(5+2\log(\frac{2r_{0}}{r^{2}(1+r)\log(1+r_{0})}))-(2^{n}(-1+n)nr^{4}(1+r^{2})\times\nonumber\\&\!&\!
\times(\frac{r_{0}}{r^{2}(1+r)\log(1+r_{0})})^{n}\log(1+r_{0})^{2}+4r_{0}^{2}(5+2\log(\frac{2r_{0}}{r^{2}(1+r)\log(1+r_{0})})))(r^{3}r_{0}^{2}(r-(1+r)\log(1+r))\log(1+r_{0})-\nonumber\\&\!&\!
-2^{1+n}(-1+n)n(-1+2n)r^{4}(1+r)^{2}(\frac{r_{0}}{r^{2}(1+r)\log(1+r_{0})})^{n}\log(1+r_{0})^{2}(-r_{0}\log(1+r)+r\log(1+r_{0}))+8r_{0}^{2}(r_{0}\times\nonumber\\&\!&\!
\times\log(1+r)-r\log(1+r_{0}))(19+6\log(\frac{2r_{0}}{r^{2}(1+r)\log(1+r_{0})}))))
\label{a16}
\end{eqnarray}
\begin{eqnarray}
&\!&\!P_{r}=\frac{1}{2r^{8}(1+r)^{2}r_{0}\log(1+r_{0})^{3}}(-2^{1+2n}(-1+n)^{2}n^{2}(-1+2n)r^{7}(1+r)^{3}\log(1+r_{0})^{4}(\frac{r_{0}}{r^{2}(1+r)\log(1+r_{0})})^{2n-1}2^{n}\times\nonumber\\&\!&\!
\times(-1+n)r^{5}(1+r)^{2}(\frac{r_{0}}{r^{2}(1+r)\log(1+r_{0})})^{n}\log(1+r_{0})^{3}(n(-1+4n)r^{3}(1+r)\log(1+r_{0})+r_{0}(16n(7+5n)-(1+2n)\times\nonumber\\&\!&\!
\times r^{3}+32n(1+n)(\frac{2r_{0}}{r^{2}(1+r)\log(1+r_{0})})))+4rr_{0}^{2}\log(1+r_{0})(-r^{3}(1+r)\log(1+r_{0})(43+14\log(\frac{2r_{0}}{r^{2}(1+r)\log(1+r_{0})}))+\nonumber\\&\!&\!
+r_{0}(-760++12r^{3}+(-544+5r^{3}-96\log(\frac{2r_{0}}{r^{2}(1+r)\log(1+r_{0})}))(\frac{2r_{0}}{r^{2}(1+r)\log(1+r_{0})})))+\log(1+r)(2^{1+2n}(-1+n)^{2}\times\nonumber\\&\!&\!
\times n^{2}(-1+2n)r^{8}(1+r)^{4}(\frac{2r_{0}}{r^{2}(1+r)\log(1+r_{0})})^{2n}\log(1+r_{0})^{4}+2r_{0}^{2}(1520r_{0}^{2}+r^{3}(1+r)\log(1+r_{0})(50r_{0}-r^{2}(1+r)\times\nonumber\\&\!&\!
\times\log(1+r_{0}))+4r_{0}(272r_{0}+3r^{3}(1+r)\log(1+r_{0}))\log(\frac{2r_{0}}{r^{2}(1+r)\log(1+r_{0})})+192r_{2}^{2}\log(\frac{2r_{0}}{r^{2}(1+r)\log(1+r_{0})})^{2})+\nonumber\\&\!&\!
+2^{n}r^{4}n(1+r)^{2}r_{0}(\frac{r_{0}}{r^{2}(1+r)\log(1+r_{0})})^{n}\log(1+r_{0})^{2}((3+4(-2+n)n)r^{3}(1+r)\log(1+r_{0})+16r_{0}(-7+2n+5n^{2}+\nonumber\\&\!&\!
+2(-1+n^{2})\log(\frac{2r_{0}}{r^{2}(1+r)\log(1+r_{0})})))))
\label{a17}
\end{eqnarray}
\begin{eqnarray}
&\!&\!P_{t}=-\frac{1}{2r^{8}(1+r)^{2}r_{0}\log(1+r_{0})^{3}}(2^{1+2n}(-1+n)^{2}n^{2}(-1+2n)r^{7}(1+r^{3}\log(1+r_{0})^{4}\log(\frac{r_{0}}{r^{2}(1+r)\log(1+r_{0})})^{2n-1}+\nonumber\\&\!&\!
+2^{-1+n}r^{5}(1+r)^{2}(\frac{r_{0}}{r^{2}(1+r)\log(1+r_{0})})^{n}\log(1+r_{0})^{3}(-2(-1+n)nr^{3}(1+r)\log(1+r_{0})+r_{0}(32(-1+n)n(7+5n)+\nonumber\\&\!&\!
+(2+n-2n^{2})r^{3}+64n(-1+n^{2})\log(\frac{2r_{0}}{r^{2}(1+r)\log(1+r_{0})})))+rr_{0}^{2}\log(1+r_{0})(r^{3}(1+r)\log(1+r_{0})(-20+r^{2}-8\times\nonumber\\&\!&\!
\times\log(\frac{2r_{0}}{r^{2}(1+r)\log(1+r_{0})}))+2r_{0}(1520-11r^{3}+4\log(\frac{2r_{0}}{r^{2}(1+r)\log(1+r_{0})})(272-r^{3}+48\times\nonumber\\&\!&\!
\times\log(\frac{2r_{0}}{r^{2}(1+r)\log(1+r_{0})}))))+\frac{1}{2}\log(1+r)(-4^{1+n}(-1+n)^{2}n^{2}(-1+2n)r^{8}(1+r)^{4}(\frac{2r_{0}}{r^{2}(1+r)\log(1+r_{0})})^{2n}\times\nonumber\\&\!&\!
\times\log(1+r_{0})^{4}+2r_{0}^{2}(-3040r_{0}^{2}+r^{3}(1+r)\log(1+r_{0})+4r_{0}(-544r_{0}+5r^{3}(1+r)\log(1+r_{0}))\times\nonumber\\&\!&\!
\times\log(\frac{2r_{0}}{r^{2}(1+r)\log(1+r_{0})})-384r_{0}^{2}\log(\frac{2r_{0}}{r^{2}(1+r)\log(1+r_{0})})^{2})+2^{n}nr^{4}(1+r)^{2}r_{0}(\frac{2r_{0}}{r^{2}(1+r)\log(1+r_{0})})^{n}\times\nonumber\\&\!&\!
\times\log(1+r_{0})^{2}((-7+6n)r^{3}(1+r)\log(1+r_{0})+32r_{0}(7-2n-5n^{2}-2(-1+n^{2})\log(\frac{2r_{0}}{r^{2}(1+r)\log(1+r_{0})})))))
\label{a18}
\end{eqnarray}

\end{document}